\documentclass[a4paper,12pt]{article}
\usepackage{psfrag}
\usepackage{graphicx}
\usepackage{graphics}
\usepackage{bm}
\usepackage{color}
\usepackage{amssymb}
\usepackage{amsmath}
\usepackage{float}

\usepackage[T1]{fontenc}
\usepackage[latin1]{inputenc}
\usepackage{amsthm}
\usepackage{braket}
\usepackage{lmodern}
\usepackage{fancyhdr} 
\usepackage{rotating}
\newcommand{\gsim}{\raise.3ex\hbox{$>$\kern-.75em\lower1ex\hbox{$\sim$}}}
\newcommand{\lsim}{\raise.3ex\hbox{$<$\kern-.75em\lower1ex\hbox{$\sim$}}}

\newcommand\hmu{{{\mu}}}

\newcommand\ba{{\mathbf{a}}}
\newcommand\bb{{\mathbf{b}}}

\newcommand\be{\begin{equation}}
\newcommand\ee{\end{equation}}
\newcommand\beq{\begin{equation}}
\newcommand\eeq{\end{equation}}
\newcommand\bea{\begin{eqnarray}}
\newcommand\eea{\end{eqnarray}}

\usepackage{vmargin}         
\setmarginsrb{2cm}{2cm}{2cm}{2cm}{0cm}{0cm}{0cm}{1cm}

\title{{\bf Gravitational wave signatures from kink proliferation on cosmic (super-) strings}}

\author{P.~Bin\'etruy\footnote{binetruy@apc.univ-paris7.fr} , A.~Boh\'e\footnote{bohe@apc.univ-paris7.fr} , T.~Hertog\footnote{hertog@apc.univ-paris7.fr} \ 
  and D.A.~Steer\footnote{steer@apc.univ-paris7.fr} \\
{\small {}}\\
{\small {\it  APC\ \footnote{Universit\'e 
Paris-Diderot, CNRS/IN2P3,  CEA/IRFU and Observatoire de Paris} ,10 rue Alice Domon et L\'eonie Duquet,
 75205 Paris Cedex 13, France}}\\
}

\begin{document}
\maketitle

\begin{abstract}

Junctions on cosmic string loops give rise to the proliferation of sharp kinks. We study the effect of this proliferation on the gravitational wave (GW) signals emitted from string networks with junctions, assuming a scaling solution.  We calculate the rate of occurrence and the distribution in amplitude of the GW bursts emitted at cusps and kinks in the frequency bands of LIGO and LISA as a function of the string tension, the number of sharp kinks on loops with junctions and the fraction of loops in the cosmological network which have junctions. Combining our results with current observational constraints, we find that pulsar data rule out a significant number of kinks on loops for strings with tensions $G\mu \gtrsim 10^{-12}$. By contrast, for smaller tensions current observations allow for a large number of kinks on loops. If this is the case, the incoherent superposition of small bursts emitted at kink-kink encounters leads to an enhanced GW background that hides the strong individual bursts from kinks and cusps.

\end{abstract}
\maketitle

\section{Introduction}

The gravitational waves (GW) emitted from cosmic string and superstring networks may enable the detection of strings with very low tensions using the LISA space interferometer \cite{Damour:2000wa}.
In this paper, following up on previous work \cite{Binetruy:2009vt,Binetruy:2010bq}, we study the effect of the presence of 3-way junctions (or Y-junctions) on the GW burst signal emitted by a cosmological network of cosmic (super-) strings. Such junctions are thought to be a generic feature of cosmic superstring networks, which contain F- and D- strings as well as bound states made up of both types. Junctions can also exist, in cosmic string networks formed in symmetry breaking phase transitions -- the simplest examples being Abelian-Higgs strings deep in the type I regime or $Z_3$ strings.

In \cite{Binetruy:2010bq} we have recently studied the dynamics of kinks interacting with Y-junctions, and found that the presence of junctions on a closed loop dramatically increases the number of large amplitude\footnote{The amplitude is defined as follows \cite{Binetruy:2010bq}. Working in Minkowski space and in the standard conformal temporal gauge, the position of a string is given by
$\textbf{x}(\sigma,t)=\frac{1}{2}\big(\textbf{a}(\sigma+t)+\textbf{b}(\sigma-t)\big)$ where $\textbf{a}^{\prime 2}=\textbf{b} ^{\prime 2}=1$. The amplitude of a discontinuity (or kink) in, say, ${\bb}'$ is defined by $A[{\bb'}] = \frac{1}{2} \| \mathbf{b}^{\prime +} -\mathbf{b}^{\prime -}\|$.}  kinks on the loop. That is, kinks proliferate. Indeed, when a kink reaches a junction on one of the strings it produces three daughter kinks --- a reflected kink and two transmitted kinks. The daughter kinks often have an amplitude (or sharpness) that is comparable to that of the incoming kink, leading to an exponentially increasing number of large amplitude kinks on the loop. As we argued in \cite{Binetruy:2010bq}, by the time proliferation ends the number of large amplitude kinks on a loop is generally several orders of magnitude larger than the number of sharp kinks expected on loops without junctions.

Here we study the effect of this proliferation on the GW burst signal emitted by string networks with junctions, focusing on models in which the different strings in the network all have similar tensions $\hmu$.
Bursts from cusps and kinks on strings give rise to high frequency GW signals that are superimposed on the low frequency spectrum of the string, and the waveforms of GW bursts on loops with junctions have been calculated in \cite{Binetruy:2009vt,Damour:2001bk}. 
In this paper, we calculate the rate of occurrence and the distribution in amplitude of the GW bursts emitted at cusps and kinks in the frequency bands of LIGO and LISA, as a function of the string tension $G\mu$, the number $k'$ of sharp kinks on loops with junctions, the fraction $q$ of loops that have junctions and the intercommutation probability $p$. To do so we will use a simple `scaling' model for the cosmic string network. We also calculate the stochastic background generated by the superposition of overlapping bursts. 

On a given loop, an individual burst emitted by a cusp is stronger than a kink burst. However, to determine the contributions of cusps and kinks to the high frequency GW burst signal in a cosmological network of cosmic strings, a more detailed analysis is required. This was done by Damour and Vilenkin in \cite{Damour:2001bk} for `standard' loops without junctions, where it was shown that bursts from cusps provide the dominant contribution to the GW signal. Here we generalize their analysis to take in account the effect of junctions.

The paper is organized as follows. In Section \ref{sec:networkevol}, we describe our scaling model for the network evolution of loops containing junctions. We compute the amplitude of bursts emitted by such a network  in Section \ref{sec:Bursts} and compare this with the sensitivity of LIGO and LISA. We then turn to the calculation of the stochastic background created by the superposition of bursts in Section \ref{stoch}, which is also affected by the kink proliferation. Finally, in Section \ref{sec:Discussion}, we combine our results from Sections \ref{sec:Bursts} and \ref{stoch} with current observational constraints and discuss the prospects for observation in the different regimes of parameter space.

\section{Network evolution}
\label{sec:networkevol}

The GW signatures from cosmic string networks depend sensitively on the details of the network evolution, and particularly on the loop distribution. Predictability is therefore still limited since, 
even in the case of standard cosmic strings with no junctions there is still significant uncertainty and disagreement on, for example, the loop distribution on the smallest scales (see, however, \cite{Lorenz:2010sm}).  However, for such networks there is broad agreement that on large scales the network reaches a scaling solution, though its dependence on the intercommutation probability $p$ of the strings is still not fully understood.  

In this paper we will follow closely the approach developed by Damour and Vilenkin in \cite{Damour:2001bk} for networks of loops without junctions. Hence we begin this section by recalling the assumptions made there before developing a generalization applicable to networks with junctions.

\subsection{Scaling solution}

Central to the GW calculations for a network of cosmic strings \emph{without} junctions (e.g.~\cite{Damour:2001bk,Olmez:2010bi,Siemens:2006vk,Siemens:2006yp}), is the fact that the network reaches a scaling solution in which
energy is lost from the network by loop formation. At time $t$ the loops are all taken \cite{Damour:2001bk} to be formed with size 
\be
\label{length}
L = \alpha t,
\ee
that is the loop production function is taken to be a $\delta$-function (see, however, \cite{Polchinski:2006ee,Rocha:2007ni,Dubath:2007mf}). There is no agreement on the value of $\alpha$, but we shall follow \cite{Damour:2001bk} and adopt the simplest scenario in which it is set by a crude estimate of the efficiency of gravitational backreaction for the damping of the small scale 
wiggles;
\begin{equation}
\label{relationalphaGmu}
\alpha=\Gamma G \mu \ , \qquad \text{ where } \qquad \Gamma=50 \ .
\end{equation}
In what follows, we keep both parameters $\alpha$ and $G\mu$ explicit in our formulae so that their origins can be traced more easily. However, when plotting our results, we will use \eqref{relationalphaGmu} to express everything in terms of $G\mu$.

Once formed, the loops decay by gravitational radiation meaning that at any given time $t$ there is a distribution of loop lengths $L$, characterised by a number density $n(L,t)$. Following \cite{Damour:2001bk} we also approximate the loop length distribution by a $\delta$-function at time $t$ (see, however, \cite{Siemens:2006vk} that goes beyond this approximation and shows it does not affect order of magnitude estimates) so that the number density of loops of size $L=\alpha t$ at time $t$ is given by
\begin{equation}
n(t)=p^{-1} \alpha^{-1} t^{-3} \ ,
\label{density}
\end{equation}
where $p\leq1$ is the reconnection probability. Notice that the typical distance between loops is approximately $n^{-1/3} = p^{1/3} \alpha^{-2/3} L \gg L$. 

We now extend this discussion to networks with junctions. The evolution of such networks has been studied in the recent literature using both numerical and analytical approaches \cite{Avgoustidis:2007aa,Hindmarsh:2006qn,Rajantie:2007hp,Tye:2005fn,Urrestilla:2007yw,Copeland:2005cy}. In \cite{Avgoustidis:2007aa} (see also \cite{Tye:2005fn}), the 1-scale model of cosmic string evolution developed in \cite{Kibble:1984hp,Martins:1996jp,Martins:2000cs}
 has been extended to string networks with junctions,
and the resulting `modified' 1-scale model equations generically have scaling solutions. One therefore expects the network with junctions to scale, and this has indeed been observed in numerical simulations \cite{Hindmarsh:2006qn,Rajantie:2007hp,Urrestilla:2007yw,Copeland:2005cy}. 

In this paper we focus on networks of strings with junctions in which the different strings in the network have similar tensions $\hmu$. In that case,
in the scaling solution the number density of loops {\it without junctions} is as given in (\ref{density}),
\begin{equation}
n(t)=p^{-1} \alpha^{-1} t^{-3} \ .
\label{densitybis}
\end{equation}
Here $p$ is again the intercommutation probability which we assume is the same for all the strings. 
On top of these loops, however, there will be a set of loops {\it with junctions} (not considered in \cite{Avgoustidis:2007aa,Tye:2005fn} but seen in simulations \cite{Urrestilla}). Consistent with the scaling solution and as we argue below, we will take the number density of such loops $n'(t)$ to be a small fraction, $q\leq 1$, of the number density of loops without junctions;
\be
n'(t) = q \cdot n(t) \,  =q \left( p^{-1} \alpha^{-1} t^{-3} \right) \, .
\label{density_junction}
\ee

\subsection{Formation of loops with junctions}

\begin{figure}[h]
\centering
   \includegraphics[scale=0.20]{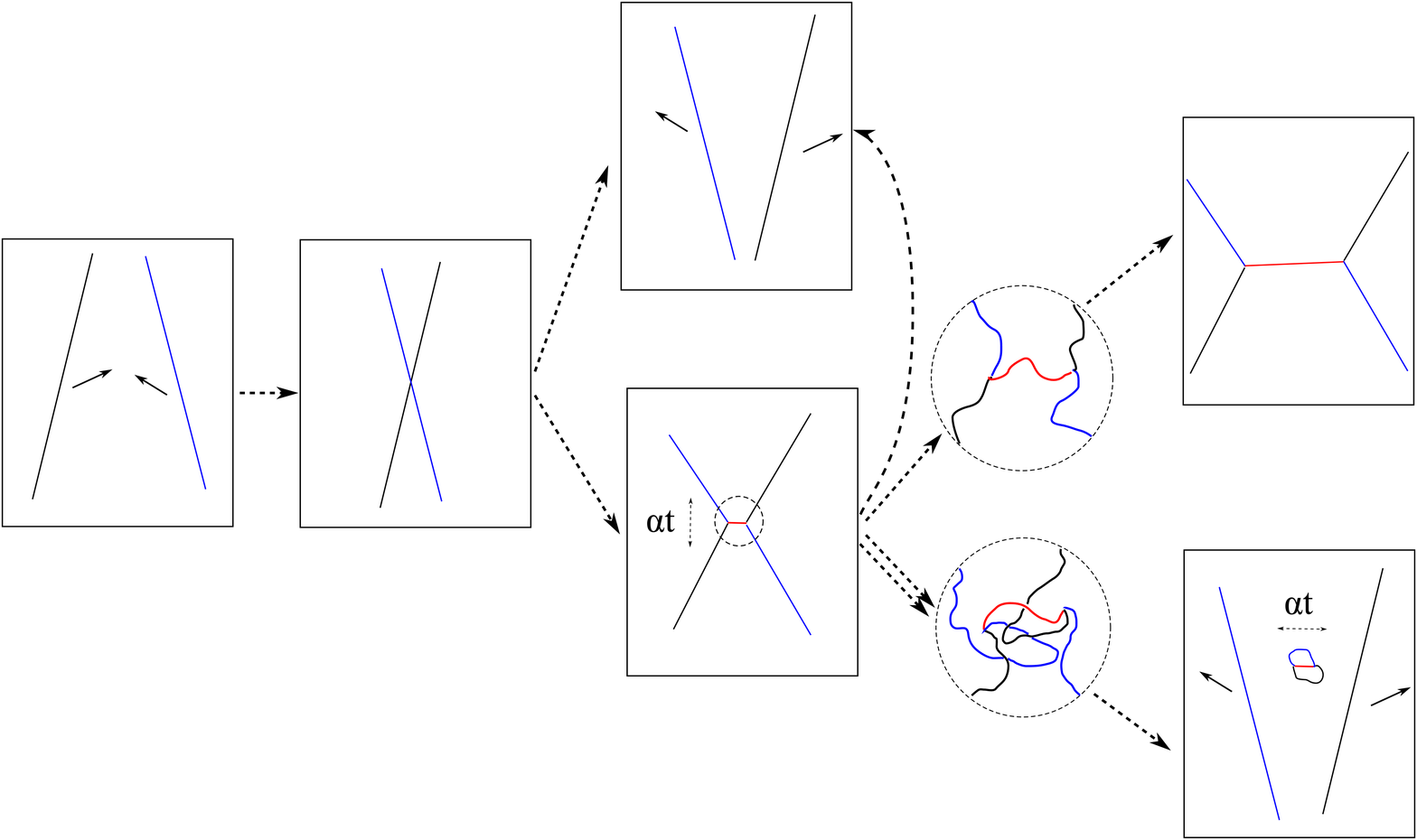} 
\caption{Formation of loops with junctions}
   \label{formationloopswithjunctions}
\end{figure}

To understand the formation of loops with junctions, and in particular the typical size of such loops, consider a network with three different strings which we differentiate by giving them a colour; say red, black and blue. A junction is a point at which these three different strings meet.  
Suppose that a string of one type
(blue on Figure \ref{formationloopswithjunctions}) collides with a string of 
another type (black). Depending on the underlying theory and on the conditions 
under which this happens (velocities, angle,...) \cite{Copeland:2006eh}, these two strings may become linked by a third string (red), thus 
generating two junctions. 
If the size of the red string remains much smaller 
than $\alpha t$, the characteristic size of the smallest wiggles, it is 
expected to shrink again to zero. By contrast, if the length of the red string becomes comparable to 
$\alpha t$, it becomes possible that 
both the blue and the black pairs of segments intersect again and form by 
exchange of partners a loop with junctions at the scale $\alpha t$. 
\footnote{if the dynamics allows for many such red links to grow much beyond 
the typical size $\alpha t$, one expects that this will lead to the formation
of a highly interconnected net, a possibility that we have discarded.}
Therefore, we assume in the following that 
all loops with junctions are produced at the same size $L=\alpha t$ as loops 
without junctions.

Since such loops can only form around junctions 
whereas standard loops can form anywhere along a string, the number of loops with junctions 
must be smaller than the number of those without junctions. In other words, we expect $q$ defined in Eq.~(\ref{density_junction}) to be smaller than 1. Furthermore, in the scaling regime where $t$ sets the only scale in the problem, we expect $q$ to be time-independent.

\section{Gravitational wave bursts}
\label{sec:Bursts}

In the previous section we have described the formation and evolution of a population of loops with and without junctions in a network of cosmic strings. In this section we determine the GW burst signal emitted from such a network and study its detectability in the LIGO and LISA frequency bands.

This amounts to a generalization of the analysis of \cite{Damour:2001bk} (to which the reader is referred for more details) that takes in account the fact that some of loops in the network contain a very large number of kinks. The different steps in this analysis are the following. One first specifies the number of kinks and cusps on an individual loop of size $L$, both with and without junctions. Then one determines the rate of GW bursts $d\dot{N}(z)$ generated by cusps and kinks from loops between redshifts $z$ and $z+dz$. The integral of this over redshift up to $z_m$ gives the total rate of burst events $\dot{N}(z_m)$ of different kinds produced by loops as a function of $z_m$. By inverting this relation and substituting this into \eqref{obssignal} one obtains the amplitude of bursts as a function of the rate. Finally we compare this with the sensitivity of the different GW interferometers and discuss to what extent these experiments will enable one to place constraints on the network parameters for the specific network model adopted here. The stochastic background generated by overlapping bursts will be discussed in section \ref{stoch}.

\subsection{Waveform}
For a burst emitted from a loop of typical size $L$ at redshift $z$, the logarithmic Fourier transform of the observed amplitude of individual GW bursts at frequency $f$ is given by the following order of magnitude estimate, 
\begin{equation}
\label{obssignal}
h(f,z) \approx \frac{G \hmu L}{((1+z) Lf)^{\beta}}\left( \frac{1+z}{t_0 z}  \right)
~\Theta(1-\theta_m(f,L,z)),
\end{equation}
where $t_0$ is the age of the universe and the angle $\theta_m$ is given by
\begin{equation}
\theta_m(f,L,z)=(f(1+z)L)^{-1/3}.
\label{angle}
\end{equation}
The Heaviside function enforces that the burst waveforms
are only valid in the high frequency domain $\theta_m <1$, and finally
the value of $\beta$ is given by
\begin{itemize}
\item $\beta=1/3$ for a burst emitted by a cusp.
\item $\beta=2/3$ for a burst emitted by a kink propagating on one of the strings.
\item $\beta=1$ for a burst produced at a kink-kink encounter or when a kink crosses a junction.
\end{itemize}

One should note that in all these three cases, numerical factors of order $1$ arising in the saddle point calculation of the waveform (see e.g.~\cite{Damour:2001bk}) have been omitted in \eqref{obssignal}. For kinks, there is an additional factor equal to the amplitude of the kink itself, defined above. In Appendix \ref{appendix:smallkinks}, however, we show that only kinks of amplitude $\mathcal{O}(1)$ contribute to the GW signal and hence we use \eqref{obssignal} in the remainder of the paper.  One should also note that the saddle point calculation of the burst waveform yields expressions containing the second derivative of the left- and right- moving waves along the string, $\ba''$ and $\bb''$, which
in Eqs.~(\ref{obssignal}) and (\ref{angle}) have been approximated by 
\be
\mathbf{a}'',\, \mathbf{b}''=\mathcal{O}(1/L)
\label{approx}
\ee
for a loop of length $L$, as in \cite{Damour:2001bk, Siemens:2006vk,Siemens:2006yp}. In this paper we consider loops with a large number $k'$ of kinks and hence one can ask if this approximation is still valid since there is now a second length-scale $L/k'$ in the problem. Despite that, we leave Eq.~(\ref{approx}) unchanged (see, however, \cite{Olmez:2010bi}) since we see no particular reason why  the curvature of the wiggles \emph{between two consecutive kinks} should differ from the usually assumed $1/L$ curvature of kinkless loops. Indeed it is straightforward to construct loops of length $L$ with numerous kinks but arbitrary curvature in between the kinks, showing that the number of kinks and the curvature scale are not in principle directly related. We therefore use Eq.~(\ref{approx}) and the corresponding expressions Eq.~(\ref{obssignal}) and (\ref{angle}). \footnote{Ultimately, however, numerical simulations of loops with many kinks must be carried out in order to determine the typical order of magnitude of $\ba'', \bb''$.}\\

\subsection{Number of cusps and kinks per loop}
\label{numbersec}

Consider first  the  population of ``standard'' periodic loops with no junctions.  
Let $c$ (resp.~$k$) denote the average number of cusps (resp.~kinks) per oscillation period $T_L=L/2$  of a loop of length $L$. Then the average number of cusps
or kinks per unit spacetime volume on such a loop is given by
\bea
\nu_{\rm cusps}&\simeq&\frac{c}{L}n(t) \, , 
\label{cuspnoj}
\\
\nu_{\rm kinks}&\simeq&\frac{k}{L}n(t)
\label{nunojunctions}
\eea
where, as in the following, we drop factors of $2$ since we are interested in order of magnitude estimates.
On loops without junctions, one typically expects $c \simeq 1$ and $k\simeq 1$.  

The situation is more complicated for the remaining loops with junctions since these 
do not evolve periodically and, as we have shown in
\cite{Binetruy:2010bq}, the number of kinks 
on loops with junctions rapidly proliferates.
This proliferation is limited, either by junction collisions, gravitational back reaction or because of the inner structure of the string.

In this paper, we assume that the typical timescale $\tau_{prolif}$ of kink proliferation is much shorter than the typical lifetime of the loop $\tau_{loop}$. Moreover, we assume that the latter is not affected by the presence of many kinks, that is to say that most of the power radiated by a loop comes from the low frequency modes so that the total power is essentially independent of the number of kinks\footnote{The validity of this assumption when the number of kink is large will be investigated in future work.}. Finally, we also assume that the typical timescale $\tau_{round~off}$ over which gravitational backreaction rounds off kinks is larger than $\tau_{loop}$ so that a kink propagates and radiates during the whole life of the loop. All these assumptions can be summarized as follows
\begin{equation}
\tau_{prolif}\ll \tau_{loop} < \tau_{round~off}
\label{timescaleassumptions}
\end{equation} 

The first inequality of \eqref{timescaleassumptions} is for instance realized if junction collisions end the proliferation process. In this case one of the strings in the loop unzips when the junctions collide, and we assume that as a result of this the loop splits into two separate loops \cite{Firouzjahi:2009nt}. These loops have no junctions and hence evolve periodically.
However, each of these contains a large (and constant) number of kinks $k' \gg 1$ propagating on the loop. The uncertainties on the exact value of $k'$ 
make it natural to consider it as a free parameter in our analysis. By analogy with the case of standard loops, we 
define $c'$ as the number of cusps per interval of time $L$:
\bea
\nu'_{\rm cusps}&\simeq&\frac{c'}{L}n'(t)=\frac{q c'}{L}n(t) 
\label{cuspjunc}
\\
\nu'_{\rm kinks}&\simeq&\frac{k'}{L}n'(t)=\frac{qk'}{L}n(t)\, .
\label{nujunctions}
\eea
Since the presence of numerous kinks is expected to inhibit the presence of cusps \cite{Garfinkle:1987yw}, we expect
\begin{equation}
c' \ll c \; (\simeq 1)\, ,
\label{cest}
\end{equation}
though our results below hold provided $c' \lsim c$.

We have estimated the order of magnitude of $k'$ in the scenario in which kink proliferation is limited by junction collision by numerically evolving loops with three strings and two junctions \cite{Binetruy:2010bq}. These simulations show that, after a typical time $\tau_{prolif} \sim 10L$, junctions collide, and at this time the total number of kinks on the loop is $\sim3^{\tau_{prolif}/L}$. However, not all kinks contribute to the signal given in Eq.~\eqref{obssignal}: as we show in Appendix A, the signal is dominated by large amplitude or `sharp' kinks.
Taking $k'$ to be the number of sharp kinks, defined as kinks with amplitude $A\gsim 0.25$, it was found \cite{Binetruy:2010bq} that each string that makes up the loop contains of order $10^3$ kinks moving in one direction (left or right). This number must be multiplied by $3$ -- the number of 
strings in the loop -- and by $2$, to account for the left moving and the right 
moving kinks, so that $k'\simeq10^4$. In cases where junctions collide later, say when $\tau_{prolif} \sim 15 L$, one easily reaches $k'\simeq 10^6$.
In fact, for more complex loops with junctions, the amplification of the 
number of kinks might continue long after the first junction collision e.g. because of the formation of new junctions \cite{Bevis:2009az}, so that values of $k'\gg10^6$ do not seem unreasonable.

In the following we concentrate on models for which the combination $qk'$ satisfies
\begin{equation}\label{rel}
qk'\gg1\, .
\end{equation}
We believe this is a realistic class of models,  since  $q \leq 1$ whereas one always expects $k'\gg1$. 
As can be seen from Eq.~(\ref{nujunctions}), from the point of view of the rate of kink events, the subnetwork of loops with junctions behaves as a network of standard loops with an effective number 
of kinks $q k'$.  Hence in the class of models we consider the average number of cusp events is dominated by loops without junctions, since $q\leq1$ and $c\gg c'$, whereas the dominant contribution to kink events comes from loops with junctions.

\subsection{Amplitude of cusp and kink bursts}
\label{ratesandamps}

From now on all quantities will be expressed in terms of redshift $z$. For simplicity we also use the same interpolating functions\footnote{In \cite{Siemens:2006vk},  it was shown that using these rather than more accurate interpolating functions (taking into account the recent change from matter to vacuum energy domination) does not affect order of magnitude estimates.}
 between $t$ and $z$ as in \cite{Damour:2001bk}. Thus
\be
\label{tfunctionofz}
t\simeq t_0 \varphi_l(z) \qquad \text{  where  } \qquad \varphi_l(z)=
(1+z)^{-3/2}(1+z/z_{eq})^{-1/2}
\ee
where $z_{eq}=10^{3.9}$ is the redshift at radiation/matter equality, and $t_0=10^{17.5}$s is the age of the universe. Below we also need $dV(z)$, the proper spatial volume\footnote{The numerical factor of $10^2$ in \eqref{vol} approximates an exact numerical factor that can be found in \cite{Damour:2001bk}.}  between redshifts $z$ and $z+dz$;
\be \label{vol}
dV(z)\simeq10^2 t_0^3 z^2(1+z)^{-13/2}(1+z/z_{eq})^{-1/2}dz \, .
\ee

Let $d\dot N_{\rm cusp/kink}(z)$ be the rate of 
bursts generated by cusps or kinks that are emitted between 
redshifts $z$ and $z+dz$ and reach us today.  Then
\bea
\label{diffratecusps}
d\dot N_{\rm cusp}(f,z)&=&\frac{\theta_m^2}{4} (1+z)^{-1} \nu_{\rm cusps}(z) dV(z)
\\
\label{diffratekinks}
d\dot N_{\rm kink}(f,z)&=&\theta_m (1+z)^{-1} \nu'_{\rm kinks}(z) dV(z)
\eea
where $\theta_m(f,z)$ is defined in Eq.~(\ref{angle}).
The first factor in each equation is the beaming fraction of the cusp/kink, namely the probability that we are
inside the set of directions of emission of a burst produced at redshift $z$.  The second factor is the usual time dilatation factor. The third factor, the number of cusps/kinks per unit spacetime volume, has been estimated in Section \ref{numbersec} and is given by
\be
\nu_{\rm cusps}(z) \simeq c \left( p^{-1} \alpha^{-2} t_0^{-4} \varphi_l(z)^{-4} \right) \, , \qquad
 \nu'_{\rm kinks}(z)\simeq qk' \left( p^{-1} \alpha^{-2} t_0^{-4} \varphi_l(z)^{-4} \right) \, .
 \label{nunujunctions}
 \ee
For each cusp or kink signal one can solve for the 
smallest redshift ($z^{\rm cusp}_m$ or $z^{\rm kink}_m$) needed to ensure a certain 
detection rate ($\dot{N}_{\rm cusp}$ or $\dot{N}_{\rm kink}$).
One should bear in mind that events occuring at a rate smaller than $1$ 
yr$^{-1}$ are not likely to be observed by experiments.  Dropping the kink/cusp label, we therefore need to solve 
 \be
 \label{integralrate}
 \dot N=\int_{0}^{z_m} d\dot N(f,z) \ ,
 \ee
This integral is dominated by the largest redshift so it follows directly from eqs.~(\ref{diffratecusps})-(\ref{nunujunctions}) that
\be \label{cusprate}
\dot N_{\rm cusp} \simeq 10^2 \; \alpha^{-8/3} (ft_0)^{-2/3} (t_0 p)^{-1} c\;  
(z_{m}^{\rm cusp})^3 (1+z_{m}^{\rm cusp})^{-7/6} (1+z_{m}^{\rm cusp}/z_{eq})^{11/6} \ , 
\ee
\be \label{kinkrate}
\dot N_{\rm kink} \simeq 10^2 \; \alpha^{-7/3} (ft_0)^{-1/3} (t_0 p)^{-1} qk'\; 
(z_{m}^{\rm kink})^3 (1+z_{m}^{\rm kink})^{-4/3} (1+z_{m}^{\rm kink}/z_{eq})^{5/3}. 
\ee

Note that in each case $\dot N$ increases monotonically with $z_m$. Inversion of the above gives $z_m = z_m (f, \dot N)$ so that, using eq.~\eqref{obssignal}, the observed signal is
\begin{equation}
\label{obssignalNnotcut}
h(f,\dot N)=h\left(f,z_m(f,\dot N)\right).
\end{equation}
The amplitude $h$ decreases with $z_m$ and thus with $\dot N$. 
Its explicit form 
reads for cusps and kinks respectively, \begin{equation}
h(f,\dot N_{\rm cusp}) \sim  \left\{ \begin{array}{ll}
	 G \mu \left(10^{-2}\dot N_{\rm cusp}t_0\right)^{-1/3} \alpha^{-2/9} (ft_0)^{-5/9} (c/p)^{1/3} \qquad \qquad \qquad (z_m^{\rm cusp} \ll 1)
 \\	
	 G \mu \left(10^{-2}\dot N_{\rm cusp}t_0\right)^{-8/11}  \alpha^{-14/11} (ft_0)^{-9/11} (c/p)^{8/11} \qquad \qquad (z_{eq} \gg z_m^{\rm cusp} \gg 1) 
	  	 \\	
	 G \mu \left(10^{-2}\dot N_{\rm cusp}t_0\right)^{-5/11}  \alpha^{-6/11} (ft_0)^{-7/11} (c/p)^{5/11} z_{eq}^{-1/2} \qquad (z_m^{\rm cusp} \gg z_{eq}) 
	\end{array}
\right.
  \label{obssignalNcusp} \\
\end{equation}
\be
h(f,\dot N_{\rm kink}) \sim  
\left\{ \begin{array}{ll}
	 G \mu \left(10^{-2}\dot N_{\rm kink}t_0\right)^{-1/3}\alpha^{-4/9} (ft_0)^{-7/9}(qk'/p)^{1/3} \qquad\qquad \qquad (z_m^{\rm kink} \ll 1) \\	
	 G \mu \left(10^{-2}\dot N_{\rm kink}t_0\right)^{-7/10}\alpha^{-13/10} (ft_0)^{-9/10}(qk'/p)^{7/10} \qquad \qquad (z_{eq} \gg z_m^{\rm kink} \gg 1) 
	 \\	
	 G \mu \left(10^{-2}\dot N_{\rm cusp}t_0\right)^{-2/5}  \alpha^{-3/5} (ft_0)^{-4/5} (qk'/p)^{2/5}  z_{eq}^{-1/2} \qquad \qquad (z_m^{\rm kink} \gg z_{eq}) 
	\end{array}
\right.
 \label{obssignalNkink}
\ee
where we have dropped the Heaviside function in eq.~(\ref{obssignal}) since it will play no role in the frequency bands and string tensions of interest.

We now argue that this by now standard procedure should be modified by the introduction of a lower redshift cutoff $z_c$.
This arises from the observation that the above analysis relies on a continuous description of the population of loops in the network (through the density $n(z)$) which, in the context of the model described in Section \ref{sec:networkevol}, is only valid above a certain 
scale (roughly $n(z)^{-1/3}$). Therefore, this approach does not apply to redshifts so small that
there are not enough loops in a sphere of the corresponding radius to allow for a statistical approach.  
Hence $z^{\rm cusp}_c$ and $z^{\rm kink}_c$ can be estimated by imposing
\begin{equation}
\label{condlargenumberinbubble}
N(z^{\rm cusp}_c)=\int_0^{z^{\rm cusp}_c} n(z)dV(z) = 1  \, , \qquad \qquad 
N(z^{\rm kink}_c)=q \int_0^{z^{\rm kink}_c} n(z)dV(z) = 1\ .
\end{equation}

The value of $z_c$ only depends on $\alpha$, $p$ and  $q$,  but not on the number 
of cusps or kinks per loop: indeed, from eqs.~(\ref{condlargenumberinbubble}), and since $z_c \ll 1$, it is straightforward to show that 
\be
z_c^{\rm cusp} \simeq \left(\frac{p\alpha}{10^2} \right)^{1/3} \, , \qquad \qquad z_c^{\rm kink} \simeq \left(\frac{p\alpha}{10^2 q} \right)^{1/3} \, .
\label{zc}
\ee
Note that since $p\leq 1$ and  $\alpha \leq 10^{-5}$, one indeed has $z^{\rm cusp}_c \ll 1$. From $z_c^{\rm kink}$ we must impose $p\alpha \ll q$ for consistency. We also emphasize that $z_c$ intrinsically depends on the density of loops in the network and not on the type of burst. The difference between cusps and kinks in \eqref{zc} is a consequence of the fact that one expects to observe cusps from the standard subnetwork (with $q=1$) and kinks from the network of loops with junctions ($q<1$).

Hence {\it if} for one or both types of burst event $z_m\ll z_c$, then the probability of there being 
one cosmic string loop inside the sphere\footnote{Note that the 
loops that we consider are basically pointlike on cosmological scales.} is $\ll 1$.
In this case, no matter how large the number of cusp/kinks per loop, 
the probability of observing anything at all is small. 
For most values of the parameters and of the rates 
considered in \cite{Damour:2001bk}, the corresponding 
redshifts $z_m$ are above the cutoff value $z_c$ so their results are
unaffected by this discussion. However, when the number of cusps or kinks per loop is very 
large, then in part of the range of rates of experimental interest 
one finds $z_m \leq z_c$ (because $z_m$ is a decreasing function of $k'$ as can be seen from \eqref{kinkrate}). Therefore, instead of \eqref{obssignalNnotcut}, one should write
\begin{equation}
\label{obssignalN}
h(f,\dot N)=h\left(f,z_m(f,\dot N)\right) \Theta\left(z_m(f,\dot N) - z_c\right)
\end{equation}
when comparing the predictions of our models with observation.

For fixed parameters ($\alpha, p, q, c, {k}'$), the lower redshift cutoff $z_c $ turns into a lower cutoff on the rate of 
events and an upper cutoff on the amplitude. Indeed, on decreasing the size of the 
sphere around oneself, the rate of events decreases and the 
amplitude of the signal increases. However, if at some point the sphere contains no loops then reducing 
its size further obviously has no effect -- the GW signal vanishes. Specifically, from eqs.~\eqref{cusprate}, \eqref{kinkrate} together with (\ref{zc}) the lower cutoff on the rates is given by
\bea
\dot N_{\rm cusp}^{\rm min}&\simeq&c \left( \frac{1}{  t_0} \right) \left( \frac{1}{\alpha^{5/3}} \right) \left( \frac{1}{(t_0 f)^{2/3}} \right)  \ ,
\label{conditioncuspsrate}
\\
\label{conditionkinksrate}
\dot N_{\rm kink}^{\rm min}& \simeq&
k'  \left( \frac{1}{  t_0} \right) \left( \frac{1}{\alpha^{4/3}} \right) \left( \frac{1}{(t_0 f)^{1/3}} \right)  \ .
\eea
Note that these are independent of $q$ and $p$. However, one sees that kink proliferation increases the minimal rate. The corresponding ($k'$-independent) cutoff on the amplitude $h(f,z_c)$ can be obtained from 
\eqref{obssignal} and is given by
\bea
h_{\rm cusp}^{\rm max}&=& 10^{2/3}\left( \frac{G \mu}{p^{1/3}}\right) \left( \frac{1}{(t_0f)^{1/3}} \right)   \alpha^{1/3} 
 \ ,
\\
\label{conditionkinksjunctionsampl}
h_{\rm kink}^{\rm max}&=& 10^{2/3} \left( \frac{G \mu}{p^{1/3}}\right) \ \left( \frac{1}{(t_0f)^{2/3}} \right) 
  q^{1/3}  \ .
\eea
In the parameter range that we will consider, $h_{\rm cusp}^{\rm max} \gg h_{\rm kink}^{\rm max}$.

\subsection{Detectability of Bursts}
\label{detect}

\subsubsection{Sensitivity of the instruments}
\label{subsec:sensitivity}

Whether or not a burst is detectable depends on the sensitivity of the experiment. We now estimate the optimal 
signal to noise ratio (SNR) of the bursts following  \cite{Damour:2001bk}. The true 
SNR $\rho$ is defined as
\begin{equation}
\rho^2=\int_0^{+\infty} \frac{df}{f^2} \frac{h_{out}(f) 
h^{*}_{temp}(f)}{S_n(f)}
\label{late}
\end{equation}
where $h_{temp}(f)$ is the (logarithmic) Fourier transform (FT) of the best 
template, $h_{out}(f)$ the (logarithmic) FT response of the 
detector and $S_n(f)$ is the (one-sided) spectral noise density. Under ideal circumstances ---
that is, when the detector output equals the actual signal arriving 
at the detector \emph{and} when the template matches exactly this signal ---
eq.~\eqref{late} reduces to
\begin{equation}
\rho^2=\int_0^{+\infty} \frac{df}{f^2} \frac{|h(f)|^2}{S_n(f)}\ .
\label{SNR}
\end{equation}
Going beyond this idealized situation requires a much more detailed analysis 
such as those performed in \cite{Siemens:2006vk,Abbott:2009rr}, but we limit ourselves to
an estimate based on \eqref{SNR} which suffices to give an idea of the detectability of the effect of kink proliferation on the GW burst signal, at least when $\rho\gg1$ and $\rho\ll1$. One should bear in mind 
however that values $\rho \sim {\cal O}(1)$ in our analysis do not allow one to draw conclusions about 
the detectability of the signal.

Using a signal waveform $h(f)=B|f|^{-\beta}e^{2\pi i f t}$, eq.~\eqref{SNR} can be rewritten as $\rho^2=|B|^2\int_0^{+\infty} \frac{df}{f} s(f)$
where $s(f)$ depends on the type of burst (the value of $\beta$) 
and on the instrument. Physically,  the value $f_c$ at which $s(f)$ reaches 
its maximum is the frequency around which the instrument is most sensitive to 
that type of burst. We can use this frequency to replace $|B|$ in the 
expression of $\rho^2$ by $h(f_c)f_c^{-\beta}$ to find an expression for the 
signal to noise ratio $\rho=h(f_c)f_c^{-\beta}(\int_0^{+\infty} \frac{df}{f} 
s(f))^{1/2}$. This can be written as
\begin{equation}
\rho=\frac{h(f_c)}{h^{\rm eff}_{n}}.
\end{equation}
Note that even though the SNR is conveniently expressed in terms of the signal at a single frequency $f_c$, its 
computation is really a noise weighted integral over the entire frequency band of the instrument. 

Since $f_c$ and $h_{n}^{\rm eff}$ do not depend much on the kind of burst, using one set of values per instrument only introduces an error of a few percent on the SNR that will not affect our order of magnitude estimates. For the sake of comparison, we adopt the values used in \cite{Damour:2001bk}, namely\\

\begin{center}
\begin{tabular}{|c|cc|}
\hline
\hline
& $f_c$ & $h_n^{\rm eff}$\\ \hline
LIGO & $150$ Hz & $1.7 \times 10^{-22}$\\ \hline
Advanced LIGO & $150$ Hz & $1.3 \times 10^{-23}$\\ \hline
LISA & $3.88~10^{-3}$Hz & $1.8 \times 10^{-22}$\\ \hline
\hline
\end{tabular}\\
\end{center}
In the figures below we plot the amplitude $h(f_c)$ and the ``SNR=1'' level 
$h^{\rm eff}_n$. Dividing the first by the latter gives an estimation of the SNR.

\subsubsection{Cusp and kink bursts in frequency band of LIGO}

\begin{figure}[H]
\centering
   \includegraphics[scale=0.40]{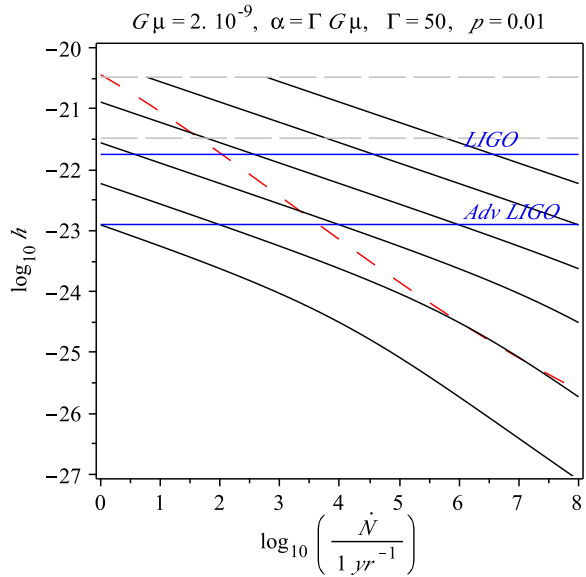} 
 \caption{
 Amplitude of the GW burst signal as a function of the rate $\dot N$ in the frequency band of LIGO/Advanced LIGO ($f_c=150 \text{Hz}$), for $p=0.01$ and $G \mu =2\times10^{-9}$. For cusps on standard loops (red dashed line) we have taken $c=1$; with these parameters $z_{eq} \gg z_m^{\rm cusp} \gg 1$ and from (\ref{obssignalNcusp}) the slope of the cusp curve is $-8/11$. For kinks (solid black lines) the different curves correspond, from bottom to top, to $q k'= 1, 10^2, 10^4, 10^6, 10^8$ and $10^{10}$.  For the large values of $qk'$, $ z_m^{\rm kink} \ll 1$ so that from  (\ref{obssignalNcusp}) the slope of the curves is $-1/3$. For $qk'=1,10$ and at large rates one also begins to enter the regime in which $z_{eq} \gg z_m^{\rm kink} \gg 1$ and the slope changes to $-7/10$. The horizontal solid blue lines correspond to the sensitivity 
(SNR = 1) of LIGO (upper curve) and Advanced LIGO (lower curve).  The horizontal dashed (grey) lines denote the maximum amplitude $h^{\rm max}_{\rm kink}$ of the kink bursts, for $q=1$ (upper curve) and $q=10^{-3}$ (lower curve). The vertical line that would indicate the confusion noise limit discussed in Section \ref{subsec:bursttobackground} lies outside the range in this figure as it corresponds to a rate $\dot{N}=150$Hz=$10^9~\text{yr}^{-1}.$}
   \label{courbesLIGOratep0p01}\end{figure}

We are now in a position to discuss our predictions for the observed GW burst amplitude $h(f, \dot{N})$, of both kinks and cusps, as a function of the observation rate $\dot{N}$ and in the frequency band of LIGO, i.e. $f=f_c=150\text{Hz}$. The burst amplitude $h(f_c, \dot{N})$ is given in \eqref{obssignalN} together with \eqref{obssignalNcusp} and \eqref{obssignalNkink}. It depends on the parameters $G\mu$ and $p$ and, in the case of kinks, also on the combination $qk'$. This combination is the fraction $q$ of loops containing junctions multiplied by the average number $k'$ of sharp kinks on loops with junctions. As mentioned above, we are particularly interested in models for which $qk' \gg 1$. Therefore to quantify the effect of kink proliferation due to the presence of junctions on the GW burst signal we plot, in Fig \ref{courbesLIGOratep0p01}, $h$ as a function of $\dot N$ for a range of different $qk'$, with fixed values of the remaining parameters $G\mu$ and $p$. We also show $h(f_c, \dot{N})$ for cusp bursts, which are not affected by the presence of junctions in our model, as well as the sensitivity limits of LIGO and Advanced LIGO.
A discussion of how observations of this kind might enable one to estimate the parameter values is given in the next subsection.

\begin{figure}[H]
\begin{minipage}{0.48\textwidth}
\centering
   \includegraphics[scale=0.40]{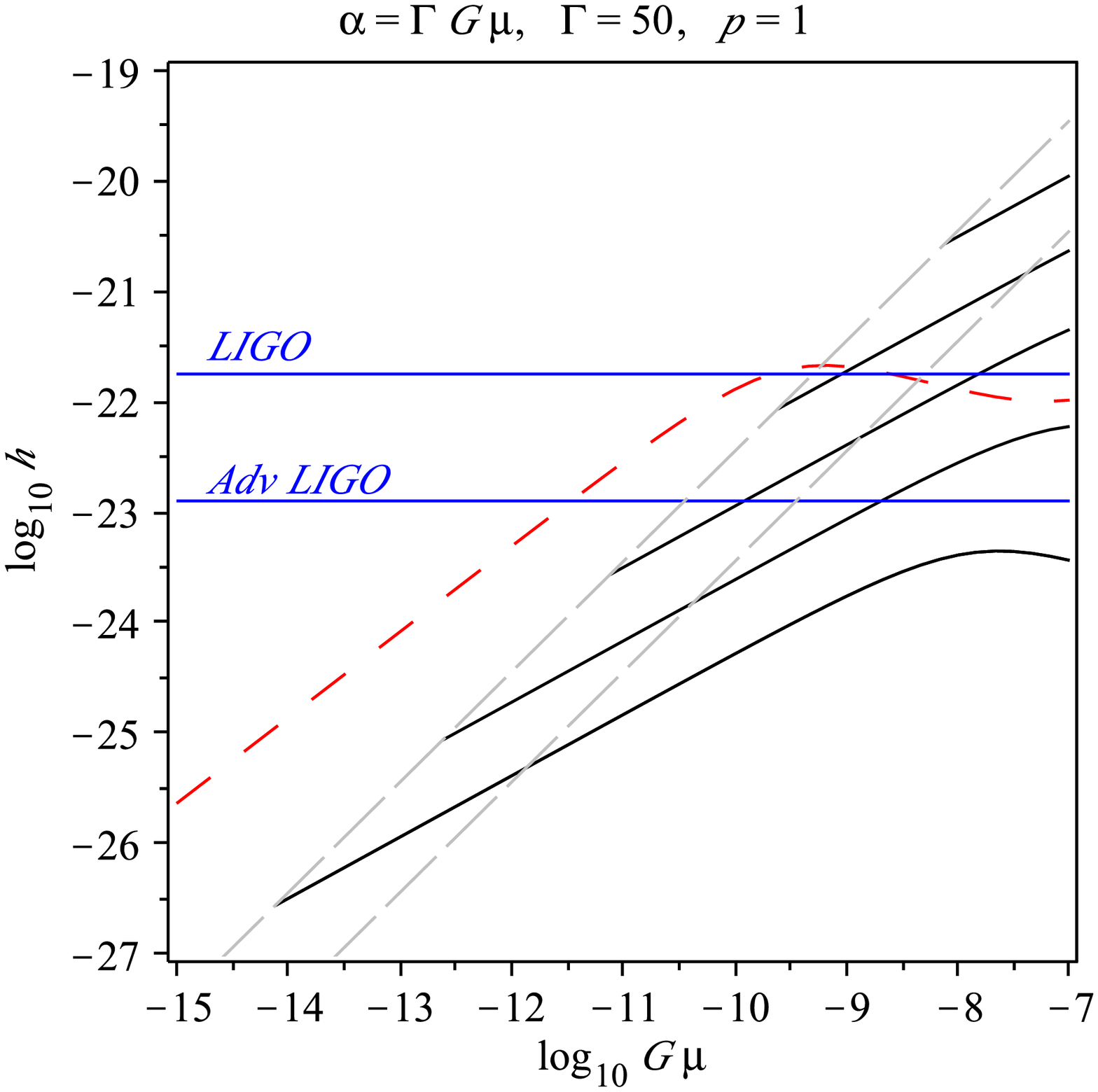} 
\end{minipage}
\begin{minipage}{0.48\textwidth}
\centering
   \includegraphics[scale=0.40]{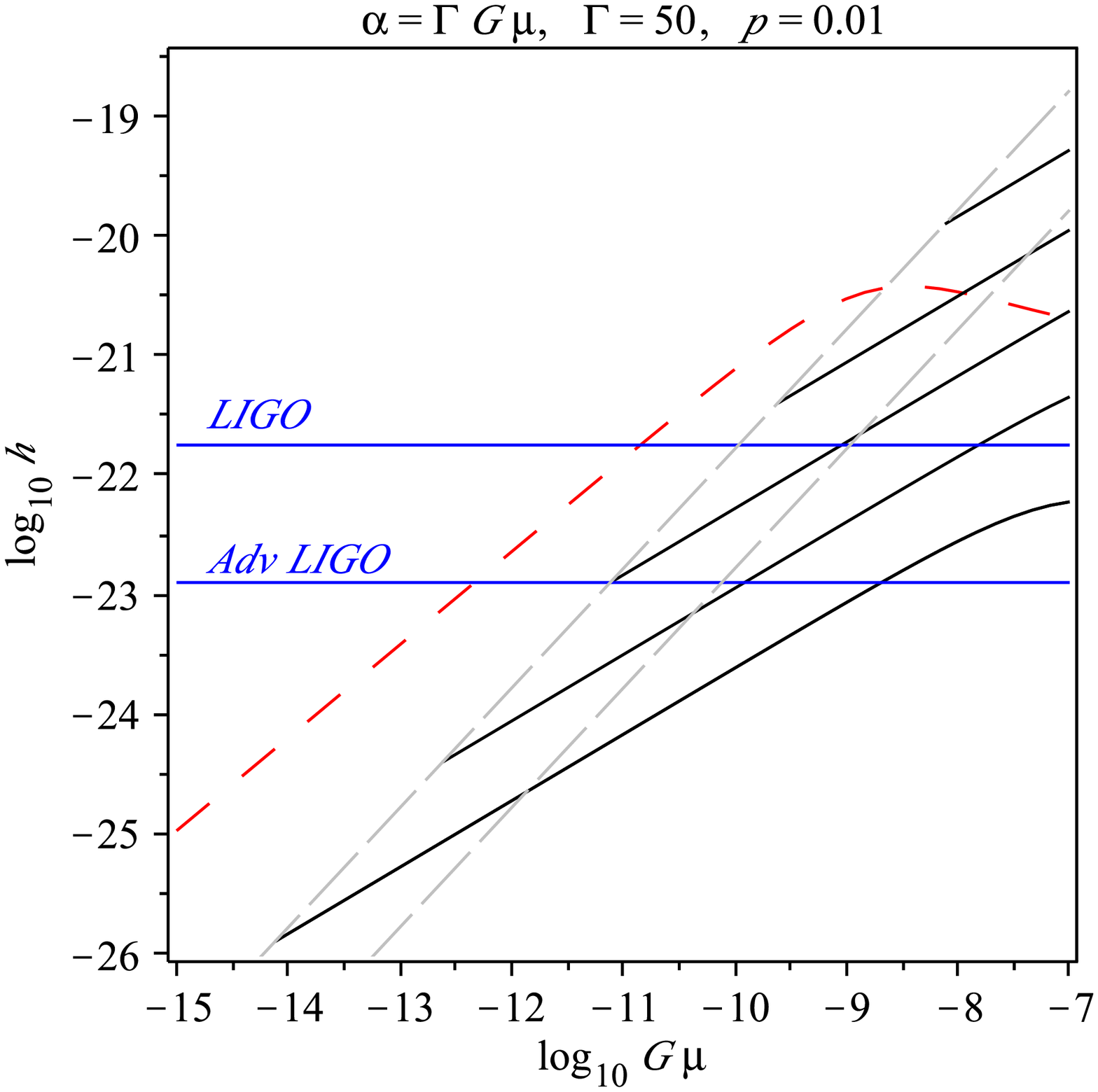} 
\end{minipage}
 \caption{
The \emph{maximum} amplitude of bursts seen by a typical observer at a random location in the LIGO/Adv LIGO frequency band during an observation time of one year, as a function of $G\mu$, for $p=1$ (left panel) and
$p=0.01$ (right panel). The red dashed curve gives the amplitude of cusps, with $c=1$. The amplitude of kink bursts is indicated by the solid black curves for (from bottom to top) $qk'=1, 10^2, 10^4, 
10^6$ and $10^8$. The kink curves have an endpoint due to the lower cutoff on the redshift.  
For smaller tensions and the same value of $qk'$, the amplitude of kink bursts is given by the grey dashed curve, which only depends on $q$ (with $q=1$ for the left curve and $q=10^{-3}$ for the right one). In this case the observation rate $\dot N$ is larger than $1$ yr$^{-1}$ as discussed in the text. 
Finally, the horizontal blue lines correspond to the sensitivity (SNR = 1) of LIGO (upper curve) and Advanced LIGO (lower curve).}
   \label{courbesLIGO}
\end{figure}

As seen in Fig. \ref{courbesLIGOratep0p01}, the proliferation of kinks on loops with junctions significantly increases the amplitude of the GW burst signal of kinks and therefore the SNR level of bursts of this kind. Even for modest values of $qk'\sim 100 $, the presence of junctions makes the GW bursts from kinks observable with LIGO when $G\mu \gsim 10^{-9}$ (see also Fig \ref{courbesLIGO} which shows the strongest bursts as a function of $G\mu$). For larger values of $qk'$  one should not only be able to observe the largest amplitude bursts from kinks, but also more distant bursts with lower amplitudes (and higher rates) hereby obtaining a curve of $h$ as a function of $\dot N$ that can be compared with $h(\dot N)$ obtained from theory.

As discussed earlier however the lower limit $z_c$ on the redshift, needed to ensure the validity of a statistical approach to interpret observations, gives rise to a $k'$-independent maximum amplitude $h_{\text{kink}}^{\text{max}}$. This is indicated by the horizontal dotted lines in Fig \ref{courbesLIGOratep0p01}, for two different values of $q$ (the upper line has $q=1$, the lower one $q=10^{-3}$). For the values of $G\mu$ and $p$ taken in Fig \ref{courbesLIGOratep0p01}, one sees that for $q=1$, $h_{\text{kink}}^{\text{max}}$ is comparable to the amplitude of the strongest (detectable) bursts from cusps and is reached for $k' \sim10^7$. 

The maximal amplitude of bursts reaching LIGO in a reasonable observation time (say with a rate larger than one burst per year) as a function of the tension is plotted in Figure \ref{courbesLIGO} for $p=1$ (left panel) and $p=0.01$ (right panel). Since the strongest bursts come from the smallest accessible redshift and have the smallest accessible rate (see Fig \ref{courbesLIGOratep0p01}), for cusps with $c=1$ (red curve), the maximal amplitude is then given by \eqref{obssignalNcusp} with $f=f_c=150 \text{ Hz }$ and $\dot{N}=1~\text{yr}^{-1}$. For kinks, the maximal amplitude is achieved either for $\dot{N}=1~\text{yr}^{-1}$ when $\dot{N}_{\rm kink}^{\rm min}<1~\text{yr}^{-1}$, i.e.~when $G\mu$ is large enough and given by \eqref{obssignalNkink} (solid black lines); or for $\dot{N}=\dot{N}_{\rm kink}^{\rm min}$ when $\dot{N}_{\rm kink}^{\rm min}>1~\text{yr}^{-1}$ for smaller tensions. In the latter case the maximal amplitude is given by \eqref{conditionkinksjunctionsampl}, which is represented in Fig \ref{courbesLIGO} by the grey dashed lines in each panel, with $q=1$ for the top curve and $q=10^{-3}$ for the bottom one. 

For small values of $G\mu \lsim 10^{-10}$, Fig \ref{courbesLIGO} shows that for all values of $k'$ cusps remain the strongest source of GW bursts that are detectable with LIGO, independently of the reconnection probability  $p$ and the fraction of loops with junctions $q$ (recall that $q\leq 1$). \emph{The presence of junctions, therefore, does not widen the range of string tensions for which cosmic strings are detectable through a GW burst of any type.} Also, even though for  $G\mu \gtrsim10^{-9}$, $h_{\rm kink}^{\text{max}}$ can exceed the amplitude of the strongest bursts from cusps (for all reasonable $p$ but provided $q$ is not too small), we will see in Section \ref{stoch} that in this regime of tensions pulsar timing observations place a rather stringent upper bound on $k'$ that essentially rules out those values for which this happens.

\subsubsection{Cusp and kink bursts in frequency band of LISA}

We now discuss the prospects for observation of the GW burst signal in the frequency band $f=f_c=3.88~10^{-3} \text{Hz}$ of LISA.

\begin{figure}[h]
\centering
   \includegraphics[scale=0.40]{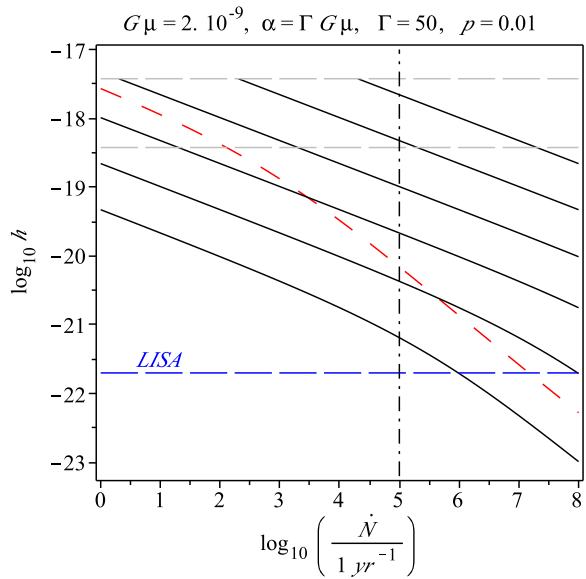} 
 \caption{Amplitude of the GW burst signal as a function of the rate $\dot N$  in the frequency band of LISA ($f_c=3.88~10^{-3} \text{Hz}$). The parameters and the meaning of the different curves are the same as in Fig \ref{courbesLIGOratep0p01} except that here $qk'=1,10^2,10^4,10^6,10^8,10^{10}$. The slope of the cusp curve changes from $-1/3$ to $-8/11$; while those of the kink curves is $-1/3$ changing, for small $qk'$ and large rates, to $-7/10$. The horizontal blue line is the "SNR=1" level of LISA given in section \ref{subsec:sensitivity}. One sees kink proliferation on loops with junctions can lead to an increase of  the SNR by several orders of magnitude, before one reaches the cutoff value that arises from the lower redshift constraint $z_c$. The vertical dashed dotted line marks the confusion noise limit $\dot{N} = 3.3 \times 10^{-3}$ Hz $= 10^5~\text{yr}^{-1}$ described in section \ref{subsec:bursttobackground} beyond which bursts overlap in the detector.}
   \label{courbesLISAratep0p01}
\end{figure}

Fig \ref{courbesLISAratep0p01} is the analogue of Fig \ref{courbesLIGOratep0p01} above. It shows the observed GW burst amplitude $h(f_c, \dot{N})$ given in Eq.~\eqref{obssignalN} as a function of the observation rate $\dot{N}$, for both kinks and cusps, and for a range of different values of $qk'$. Apart from the value of the frequency, the parameters are identical to those of Fig \ref{courbesLIGOratep0p01}. The fact that  the burst amplitude is inversely proportional to a power of the frequency (see Eqs.~\eqref{obssignalNcusp}-\eqref{obssignalNkink}) means the amplitude of the signal in the LISA band is much larger compared to the LIGO band. This leads to a much larger SNR, so that observations of GW bursts in the LISA band open up a new window in parameter space in which cosmic string networks are detectable through their gravitational wave burst signal. As before, the lower redshift limit $z_c$ leads to a maximum amplitude $h_{\rm kink}^{\text{max}}$ which, for $G\mu \sim 10^{-9}$, is reached for $k' \sim 10^5$. As can be seen in Fig \ref{courbesLISAratep0p01}, even larger values of $k'$ lead to a wide range of high burst rates in which kinks provide the dominant contribution to the GW burst signal provided $q\approx1$.

\begin{figure}[H]
\centering
\begin{minipage}{0.48\textwidth}
\centering
   \includegraphics[scale=0.40]{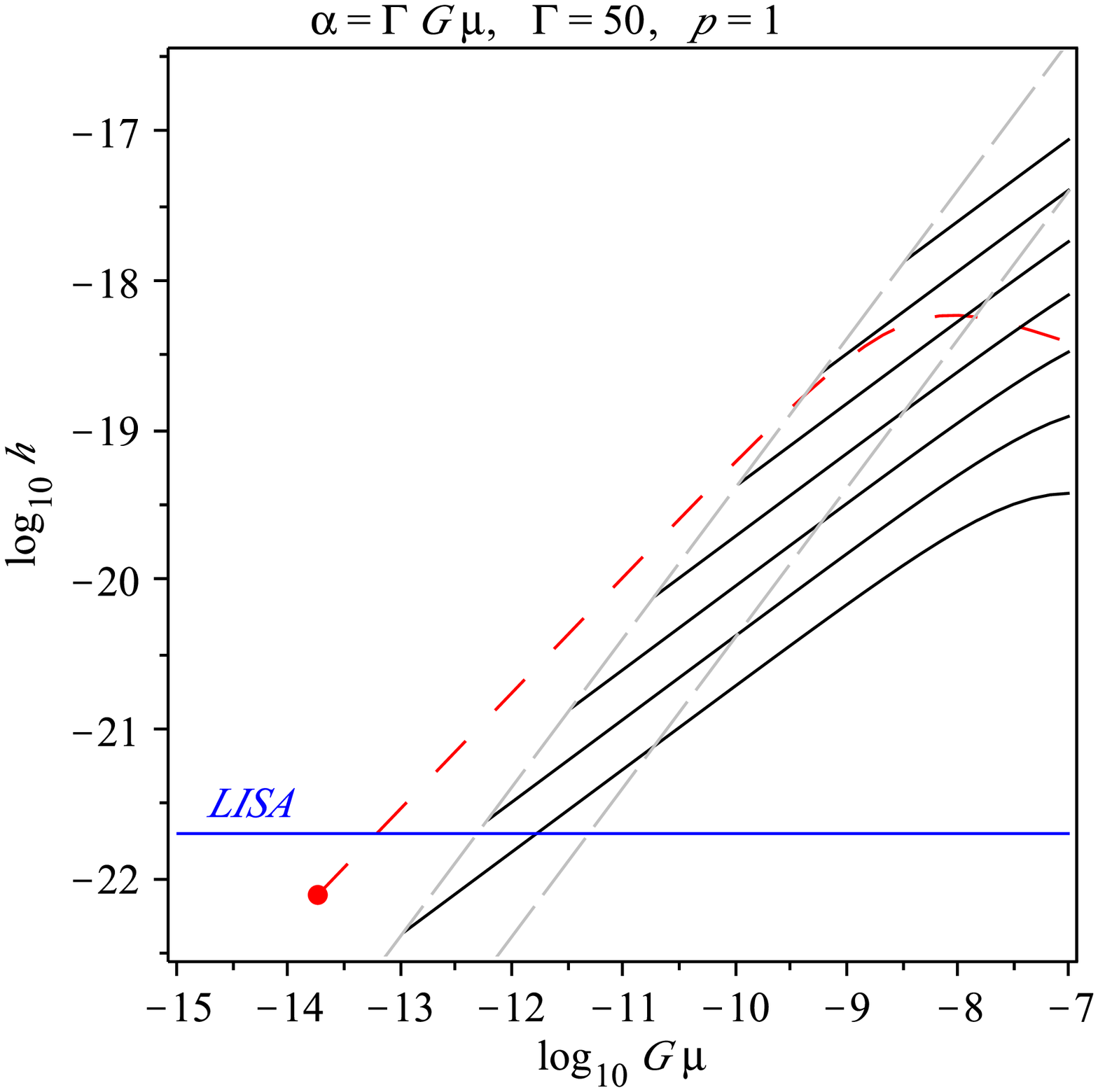} 
\end{minipage}
\begin{minipage}{0.48\textwidth}
\centering
   \includegraphics[scale=0.40]{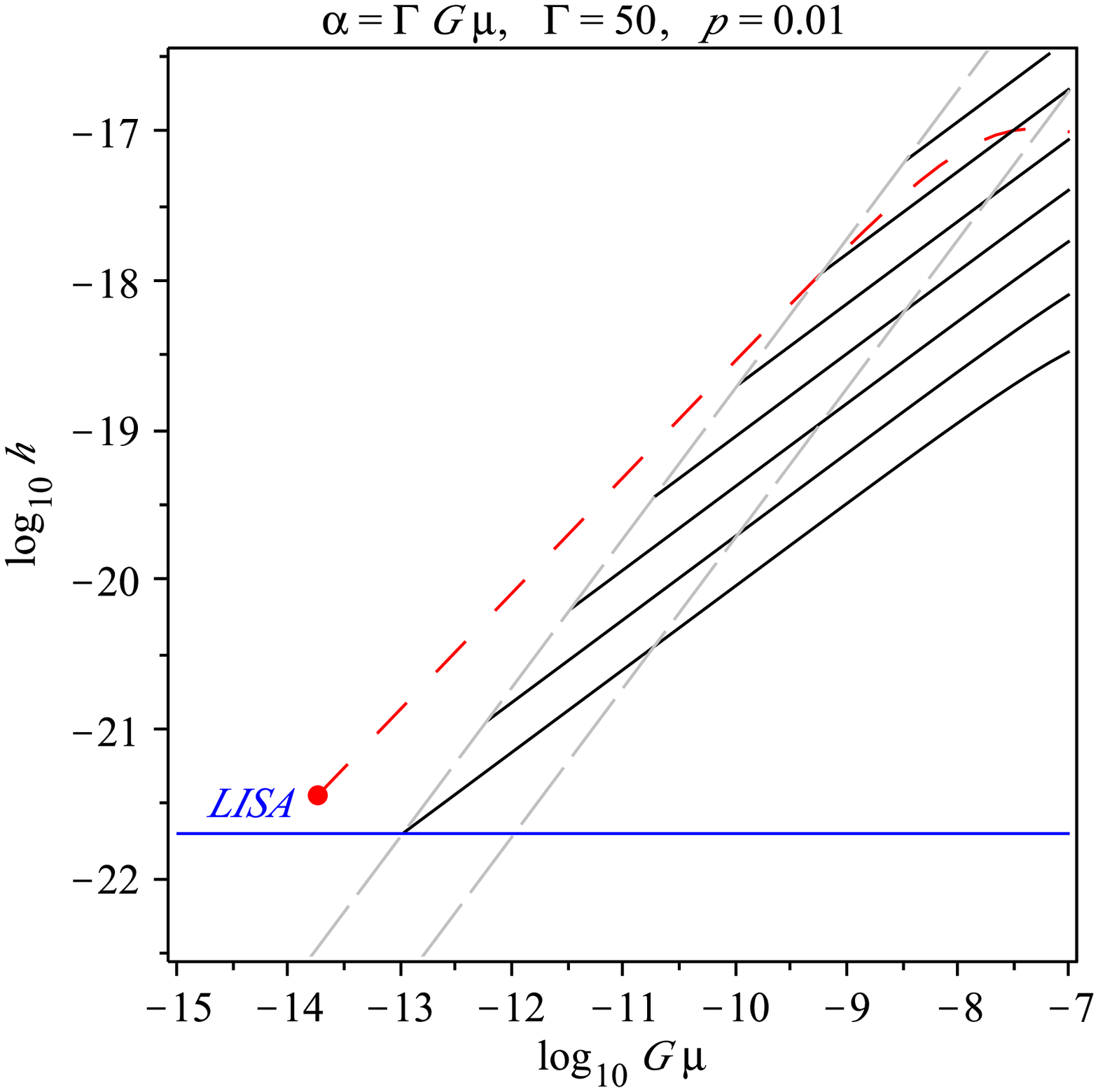} 
\end{minipage}
 \caption{
The maximum amplitude of bursts seen by a typical observer at a random location in the LISA band during an observation time of one year, as a function of $G\mu$, for $p=1$ (left panel) and
$p=0.01$ (right panel). The red dashed curve gives the amplitude of cusps, with $c=1$. The amplitude of kink bursts is indicated by the solid black curves for (from bottom to top) $qk'=1, 10, 10^2, 10^3, 10^4, 10^5$ and $10^6$. Both the cusp and the kink curves have an endpoint due to the lower cutoff on the redshift.  
For smaller tensions and the same value of $qk'$, the amplitude of bursts is bounded by the grey dashed curve, which only depends on $q$ (with $q=1$ for the left curve and $q=10^{-3}$ for the right one). In this case the observation rate $\dot N$ is larger than 1/year as discussed in the text. 
Finally, the horizontal blue line corresponds to the LISA sensitivity (SNR = 1).}
   \label{courbesLISAratep1}
\end{figure}

Fig \ref{courbesLISAratep1} in turn is the analogue of Fig \ref{courbesLIGO}: it shows the maximum amplitude of bursts reaching the interferometer during a reasonable observation time, at a rate
$\dot N \ge 1\ \text{yr}^{-1}$, as a function of $G\mu$ and for a range of different $qk'$  (note though that here the kink curves are for $qk'=1,10,10^2,\ldots,10^6$). One sees that kinks on standard loops without junctions, namely $qk'=1$, should be detectable with LISA for tensions $\gtrsim 10^{-12}$. The presence of junctions does not significantly broaden this range. However, in the region where kinks are detectable, the proliferation of kinks on loops with junctions leads almost certainly to an increase of  the SNR by several orders of magnitude, before reaching its cutoff value that arises from the lower redshift constraint $z_c$. We also note that the cusp curve in Fig \ref{courbesLISAratep1} has an endpoint, at the ($p$-independent) value $G\mu\approx[c^{-1}t_0(t_0f)^{2/3} (1\text{yr}^{-1})]^{-3/5}$ determined from \eqref{conditioncuspsrate}. This has a similar origin as the endpoints of the kink curves discussed above: As $G\mu$ decreases, $z_m$ for cusps goes below $z_c$ since $z_m \sim (G\mu)^{8/9}$ whereas  $z_c \sim (G\mu)^{1/3}$, hereby cutting off the observed curve.

The large burst amplitudes generally found in the LISA band means there is a priori a rather wide regime in which observations might enable one to determine the values of some or even all of the four parameters  $G\mu$, $p$, $q$ and $k'$ in our problem. (We will see in Section 4, however, that the stochastic background constrains the window in which this is actually possible.) To estimate the parameters one ought to combine the information\footnote{We make no use of information on the position on the sky of the sources of the detected bursts \cite{Cohen:2010xd}.} from the amplitudes of the observed kink and cusp bursts with the additional constraint(s) obtained from one (or two) endpoints of the $h(\dot N)$ curves or from changes in the slope of $h(\dot N)$. The latter approach is possible only for $G\mu \gtrsim 10^{-9}$ (see Fig \ref{courbesLISAratep0p01}). For $G\mu \lesssim 10^{-10}$ we find the observable bursts are all emitted at small redshift, so that the second derivative of $h(\dot N)$ essentially vanishes, yielding no information. However, in this regime the coordinates of the endpoint(s) of the kink and/or cusps curves in the $(\dot N, h)$-plane are observable, at least for some values of the parameters, which provides one or two additional relation(s) between the parameters (on top of the overall amplitude of the curves).

\subsection{Other bursts}
\label{other}

So far we have only discussed bursts generated by kinks and cusps.  We now comment on bursts produced at  kink-kink encounters (k-k bursts) and when kinks cross junctions (k-j). 

The waveforms of these were calculated in \cite{Binetruy:2009vt} and are given in \eqref{obssignal}.
For a given frequency and redshift these bursts have a lower amplitude than those emitted by cusps and kinks. However, this could be compensated by the fact that they are emitted in all directions in space. 
Let us start with kink-kink bursts. Following the same derivation as in subsection \ref{ratesandamps},
\bea
\label{diffratekk}
d\dot N_{\rm k-k}(f,z)&=&(1+z)^{-1} \nu'_{\rm k-k}(z) dV(z)
\eea
where
\bea
\nu'_{\rm k-k}(z) &\simeq& q k'^2 \left( p^{-1} \alpha^{-2} t_0^{-4} \varphi_l(z)^{-4} \right).
 \label{nukkkj}
\eea
In analogy with (\ref{conditionkinksjunctionsampl}), one can show that the maximum cutoff on the amplitude of k-k bursts on loops with junctions is given by
\begin{equation}
h_{{\rm k-k}}^{\rm max}=G\mu\ \alpha^{-1/3}(ft_0)^{-1}p^{-1/3}q^{1/3}.
\end{equation}
For LIGO this is below the instrument sensitivity in all parameter space (the requirement for individual kink-kink bursts  to be observable is $(G\mu)^2 q/p>10^{-5},10^{-8}$ for LIGO, Advanced LIGO). For LISA the condition is $(G\mu)^2 q/p > 10^{-18}$, so that kink-kink bursts should be observable provided $G\mu$ and $q$ are not too small. The corresponding cutoff on the k-k rate is
\begin{equation}
\dot N_{\rm k-k}^{\rm min}=k'^2(\alpha t_0)^{-1}
\end{equation}
which can be very large since it is proportional to $k'^2$. In fact, when $k'^2(G\mu)^{-1}>10^{21}$ (resp. $10^{16}$) in the LIGO (resp.~LISA) band, this will be close to or above the confusion noise limit discussed below in section \ref{subsec:bursttobackground}. This implies that individual kink-kink bursts should not be observable: the sensitivity of LIGO is not sufficient, and whereas the sensitivity of LISA allows for their observation, the bursts are in the confusion noise region. However, as we show in Section \ref{stoch}, the superposition of many bursts of this kind leads to a significant enhancement of the GW stochastic background.

It is more difficult to estimate the rate of kink-junction events but because kink-junction interactions are far less frequent than kink-kink interactions, one expects the observed rate of k-j bursts  to be negligibly small compared to the rate of k-k bursts. Therefore, since both types of bursts have the same waveform, it seems plausible that k-j bursts are irrelevant from an observational point of view.

\subsection{From individual bursts to a stochastic background}

\label{subsec:bursttobackground}

Bursts that are produced at large redshifts have small amplitudes but large rates. Therefore they can overlap at the detector where their superposition shows up as a Gaussian stochastic background. In \cite{Damour:2001bk} it was shown that the bursts contributing to the stochastic background seen by an instrument observing at a typical frequency $ f_c$  are those for which the rate satisfies
\begin{equation}
\label{cn}
\dot{N}> f_c.
\end{equation}
This bound arises from the fact that, given the burst waveforms, the duration $\tau$ of a burst in the time domain is of the order of their lower frequency cutoff, imposed by $\Theta(1-\theta_m(f,L,z))$ \cite{Siemens:2006vk}. However, since interferometers have larger lower cutoff frequencies\footnote{An order of magnitude estimate of the cutoff that each instrument applies to different kinds of bursts is given by $f_c$, defined in section \ref{subsec:sensitivity}}, the effective duration of bursts in the detectors is in fact $\tau\approx f_c^{-1}$. During an interval of $\tau$, the number of bursts that arrive at the detector is of the order of $\dot{N}\tau$. Therefore one expects to observe a superposition of bursts instead of individual bursts when $\dot{N}> f_c$.

In our plots of the burst amplitude versus the rate, the inequality \eqref{cn} translates into a vertical line. Individually detectable bursts must lie to the left of these vertical lines\footnote{As we discuss in Section \ref{stoch}, the existence of a stochastic background has implications for the detectability of individual bursts even to the left of this line, since it acts as a "self confusion" noise that could hide the bursts.}. In the frequency band of LISA the limit is shown in Fig \ref{courbesLISAratep0p01}, where the confusion noise area is given by $\dot{N}>10^5$/year. In the frequency band of LIGO/Advanced LIGO, the limiting rate is $\dot{N}>10^9$/year so that the vertical line lies outside the range of $\dot N$ in Fig \ref{courbesLIGOratep0p01}. The next section is devoted to the computation of the characteristic amplitude of the stochastic background.
 
\section{Superposition of bursts}
\label{stoch}

\subsection{Characteristic amplitude of stochastic background}

Bursts for which the rate satisfies \eqref{cn} overlap at the detector and show up as a stochastic background rather than as individual bursts. A stochastic background of gravitational waves is conventionally characterized by the spectral energy density of gravitational waves,
\begin{equation}
\Omega_{gw}(f)=\frac{1}{\rho_c}\frac{d \rho_{gw}}{d \ln f}
\end{equation}
where $\rho _{gw}$ is the energy density in GWs and $\rho_c$ the critical density of the universe.  For our discussion it will be useful to introduce the
(dimensionless) characteristic (strain) amplitude of the background, $h_c(f)$, defined as
\begin{equation}
\langle h^*(f)h(f') \rangle =f~\delta(f-f')~h^2_c(f)
\end{equation}
where, as in the remainder of this paper, $h(f)$ is the (dimensionless) logarithmic FT of the strain. In terms of $h_c(f)$, one finds $\Omega_{gw}(f)$ is given by
\begin{equation}
\Omega_{gw}(f)=\frac{3\pi^2}{2}(ft_0)^2 h_c^2(f).
\label{omegatohc}
\end{equation}

Following \cite{Damour:2001bk}, the characteristic amplitude $h_c(f)$ of the background generated by the superposition of bursts is obtained by integrating the square amplitude of individual bursts emitted at redshift $z$, weighted by the number of bursts emitted at that redshift:
\be
\label{intbackground}
h_c^2(f)=\int_{{\rm max}(z_c,z_{b\rightarrow b})}^{z_{hf}}  h^2( f,z) \frac{\dot{N}(f,z)}{f}\frac{dz}{z}
\ee
where $\dot{N}(f,z)$ is given in (\ref{cusprate}) and (\ref{kinkrate}) for cusps and kinks respectively.
The lower bound on the integral is $z_{b\rightarrow b}$, where the subscript stands for `bursts to background', which is the minimal redshift for which bursts arrive superimposed. It follows from \eqref{cn} that $z_{b\rightarrow b}$ is the solution of
\begin{equation}
\dot{N}(z_{b\rightarrow b})=f,
\label{zbb}
\end{equation}
provided of course that this is larger than $z_c$. Note that for bursts emitted by kinks, $z_{b\rightarrow b}$ depends on $f$, $\alpha$, $p$, $qk'$. 

The integral \eqref{intbackground} must be restricted to values of $z$ satisfying $\theta_m(z)<1$, where $\theta_m$ is given in \eqref{angle}, to ensure it represents the sum of high frequency bursts. Since $\theta_m(z)$ is an increasing function of $z$, this leads to an upper (high frequency) bound $z_{hf}$ for the integral given by
\begin{equation}
\theta_m(z_{hf})=1.
\end{equation} 
By inserting \eqref{tfunctionofz} in \eqref{length} and then inverting \eqref{angle}, one easily obtains the smooth interpolating approximation 
\begin{equation}
z_{hf}(\alpha,f)=\Theta\left(f-\frac{1}{t_0 \alpha}\right)\left(1+ft_0\alpha\right)^2\left(1+\frac{\alpha t_0 f}{z_{eq}^{1/2}}\right)^{-1}.
\label{zhfanalytic}
\end{equation}
The Heaviside function implies that $z_{hf} \rightarrow 0$ at $f_{cut}=\frac{1}{\alpha t_0}$. Hence, for a given value of $\alpha$ all spectra are cut at $f_{cut}$ regardless of the type of burst involved.

Until recently, only the cusp background had been calculated under the assumption that this was the leading background. Equation 4.8 in \cite{Damour:2004kw} gives an analytic approximation to $h_c^{\rm cusp}$ valid for small frequencies. With our choice of parameters this reads
\begin{equation}
h_c^{\rm cusp}\approx10 (G\mu) \alpha^{-2/3}p^{-1/2}(ft_0)^{-7/6}.
\end{equation}
Recently, however, it was shown \cite{Olmez:2010bi} that the amplitude of the kink background is comparable to that of the cusp one, even with the standard assumption that the number of kinks per loop is of order $1$. This calculation relies on a numerical computation of the integral appearing in \eqref{intbackground} as well as on an analytical approximation of both contributions.

In the next subsection, we approximate analytically for all frequencies the kink and kink-kink backgrounds and we show that even for modest values of $k'$, the latter gives the dominant contribution to the GW background.

\subsection{Different contributions to GW background}

\subsubsection{The kink contribution}

For kink bursts, using eqs~
(\ref{obssignal}) and (\ref{kinkrate}), the characteristic strain \eqref{intbackground} reduces to
\begin{equation}
h_c^{\rm kink}= \sqrt{10^2}(G \mu) p^{-1/2} \alpha^{-5/6} (t_0 f)^{-4/3} (qk')^{1/2} \left( \int _{max(z_c^{\rm kink},z_{b\rightarrow b}^{\rm kink})} ^{z_{hf}}  \frac{(1+z/z_{eq})^{4/3}}{(1+z)^{5/3}}dz\right)^{1/2}.
\label{hckinkintegral}
\end{equation}

We show in Appendix \ref{appendix:integrals} that in the regime in which $z_{hf}\gg z_{eq}^2$ the integral is not sensitive to variations of the lower bounds\footnote{One should stress that this is very different from the cusp case studied in \cite{Damour:2001bk}. One of the important observations of that work was that the calculation of $h_c$ had to be done by integrating from $z_{b\rightarrow b}$ (instead of from $z=0$) in order to avoid counting non overlaping bursts as part of the stochastic background. Though this remains obviously true for the kink (and kink-kink) background, our study shows that integrating from $z=0$ would only introduce a negligible error in the computation of $h_c^{k/k-k}$. This is because at large frequencies, $h_c^{k/k-k}$ is dominated by bursts produced around $z_{hf}$ and at low frequencies by bursts produced around $z=1$ so the exact value of $z_{b\rightarrow b}$ does not matter. By contrast, in the case of cusps, starting the integral at $z=0$ does introduce an important error in certain regimes. 
} and scales as $z_{hf}^{2/3}/z_{eq}^{4/3}$. Using \eqref{zhfanalytic}, we see that the condition $z_{hf}\gg z_{eq}^2$ translates into $\alpha t_0 f \gg z_{eq}^{3/2}$.
This yields
\begin{equation}
h_c^{\rm kink}\approx 10\ z_{eq}^{-1/2} (G \mu) \alpha^{-1/2} p^{-1/2} (qk')^{1/2} (t_0 f)^{-1}\qquad  \qquad \text{for } f > \frac{z_{eq}^{3/2}}{\alpha t_0}
\label{hckinklargef}
\end{equation}
Inserting this into \eqref{omegatohc} shows that the spectrum is flat at large frequencies and the amplitude scales as $G\mu/p$ (see also equation (63) of \cite{Olmez:2010bi}).

We now need to study the regime in which $1\ll \alpha t_0 f \ll z_{eq}^{3/2}$ or, in terms of $z_{hf}$, $1\ll z_{hf}\ll z_{eq}^2$. The key point here is that both $z_c^{\rm kink}$ and $z_{b\rightarrow b}^{\rm kink}$ will be in the $z\ll1$ region. This was already discussed in section \ref{ratesandamps} for $z_c^{\rm kink}$ (and is true provided that $q$ is not too small). Using \eqref{zbb} and \eqref{kinkrate}, we see that the condition $z_{b\rightarrow b}^{\rm kink}\ll1$ translates into $(\alpha t_0 f)^{4/3}\ll10^2(qk')\alpha^{-1}p^{-1}$ and since we are interested in $\alpha t_0 f \ll z_{eq}^{3/2}$, a necessary condition for $z_{b\rightarrow b}^{\rm kink}$ to be $\ll1$ is $10^6\approx10^{-2}z_{eq}^{2}\ll (qk')\alpha^{-1}p^{-1}$ which is satisfied for all values of the parameters. We are therefore in the situation described in the appendix where the main contribution to the integral comes from $z\approx1$ and the value of the integral is independent of the bounds. This then trivially leads to 
\begin{equation}
h_c^{\rm kink}= 10 (G \mu) p^{-1/2} \alpha^{-5/6} (t_0 f)^{-4/3} (qk')^{1/2}\qquad  \qquad \text{for } \frac{1}{\alpha t_0}< f < \frac{z_{eq}^{3/2}}{\alpha t_0}.
\label{hckinksmallf}
\end{equation}
It will be useful to write an interpolating function approximating $h_c^{\rm kink}$ at all frequencies
\begin{equation}
h_c^{\rm kink}\approx 10 (G\mu) \alpha^{1/2} p^{-1/2} (qk')^{1/2} \Theta\left(f-\frac{1}{\alpha t_0}\right)\left(1+\alpha t_0 f\right)^{-4/3} \left(1+\frac{\alpha t_0 f}{z_{eq}^{3/2}}\right)^{1/3} \, .
\label{hckinkinterpolating}
\end{equation}

An important observation is that $h_c^{\rm kink}$ is proportional to $k'^{1/2}$ so that, unlike in the case of bursts, arbitrarily long kink proliferation would make the amplitude of the background arbitrarily large.

\subsubsection{The kink-kink (kk) contribution}

Kink-kink bursts have been discussed in section \ref{other}. In this case
$h_c$ is given by
\begin{equation}
h_c^{\rm k-k}= \sqrt{10^2} (G\mu) \alpha^{-1} p^{-1/2} (t_0f)^{-3/2} (qk'^2)^{1/2}\left( \int _{\max(z_c,z_{b\rightarrow b}^{\rm k-k})} ^{z_{hf}} \left(\frac{1+z/z_{eq}}{1+z}\right)^{3/2}dz\right)^{1/2}\, .
\label{hckkintegral}
\end{equation}
The derivation of an analytic expression for $h_c^{\rm k-k}$ proceeds in a similar way as the kink case above.

Following the discussion in Appendix \ref{appendix:integrals}, we first note that in the regime in which $z_{hf}\gg z_{eq}^{3/2}$, the integral is independent of the value of the lower bound and equal to $z_{hf}/z_{eq}^{3/2}=z_{eq}^{-1}(\alpha t_0 f)$. This leads to
\begin{equation}
h_c^{\rm k-k}\approx  10 z_{eq}^{-1/2} (G\mu) \alpha^{-1/2} p^{-1/2}(qk'^2)^{1/2} (t_0 f)^{-1}\qquad  \qquad \text{for } f > \frac{z_{eq}}{\alpha t_0}.
\label{hckklargef}
\end{equation}
As in the kink case, $\Omega_{gf}$ is flat at large frequencies.
When $1\ll \alpha t_0 f \ll z_{eq}$, both $z_c^{\rm k-k}$ (which is equal to $z_c^{\rm kink}$ because it only depends on the population of loops with junctions) and $z_{b\rightarrow b}^{\rm k-k}$
are $\ll1$ so that \eqref{hckklargef} is dominated by $z\approx 1$ and is of order $1$. This yields
\begin{equation}
h_c^{\rm k-k}\approx  10 (G\mu) \alpha^{1/2} p^{-1/2}(qk'^2)^{1/2} (\alpha t_0 f)^{-3/2}\qquad  \qquad \text{for }\frac{1}{\alpha t_0}< f < \frac{z_{eq}}{\alpha t_0}\, .
\label{hckksmallf}
\end{equation}
Once again, we can summarize our results using a smooth interpolating function
\begin{equation}
h_c^{\rm k-k}\approx  10 (G\mu) \alpha^{1/2} p^{-1/2} (qk'^2)^{1/2} \Theta\left(f-\frac{1}{\alpha t_0}\right)\left(1+\alpha t_0 f\right)^{-3/2} \left(1+\frac{\alpha t_0 f}{z_{eq}}\right)^{1/2} \, .
\label{hckkinterpolating}
\end{equation}
Note that increasing $qk'^2$ simply increases the amplitude of the spectrum whilst leaving its frequency dependence unchanged. Decreasing $\alpha$ means decreasing the amplitude while shifting the spectrum to higher frequencies.

\subsubsection{Comparison with cusp background}
Dividing \eqref{hckkinterpolating} by \eqref{hckinkinterpolating}, one obtains $h_c^{\rm k-k}/h_c^{\rm kink}=\sqrt{k'} \Phi(\alpha t_0 f)$ where $\Phi(x)=(1+x)^{1/6}(1+x/z_{eq})^{-1/2}(1+x/z_{eq}^{3/2})^{1/3}$. This function is of order $1$ for all values of $x$ so we conclude that $h_c^{\rm k-k}$ is always larger than $h_c^{k}$ by a factor of order $\sqrt{k'}$. Note that \emph{this also implies that in the absence of kink proliferation, when $q=k'=1$, kink-kink events contribute at the same order as kink bursts to the stochastic background.}

Furthermore, it was shown in \cite{Olmez:2010bi} that the contribution from cusps to the stochastic background is of comparable order to the standard kink one ($q=k'=1$). \emph{We therefore conclude that as soon as $qk'\gg1$, the dominant contribution to the GW background comes from the k-k bursts. For this reason, from now on we mainly focus on the k-k background.}

\subsection{Self-confusion noise}

\begin{figure}[H]
\begin{minipage}{0.48\textwidth}
\centering
   \includegraphics[scale=0.40]{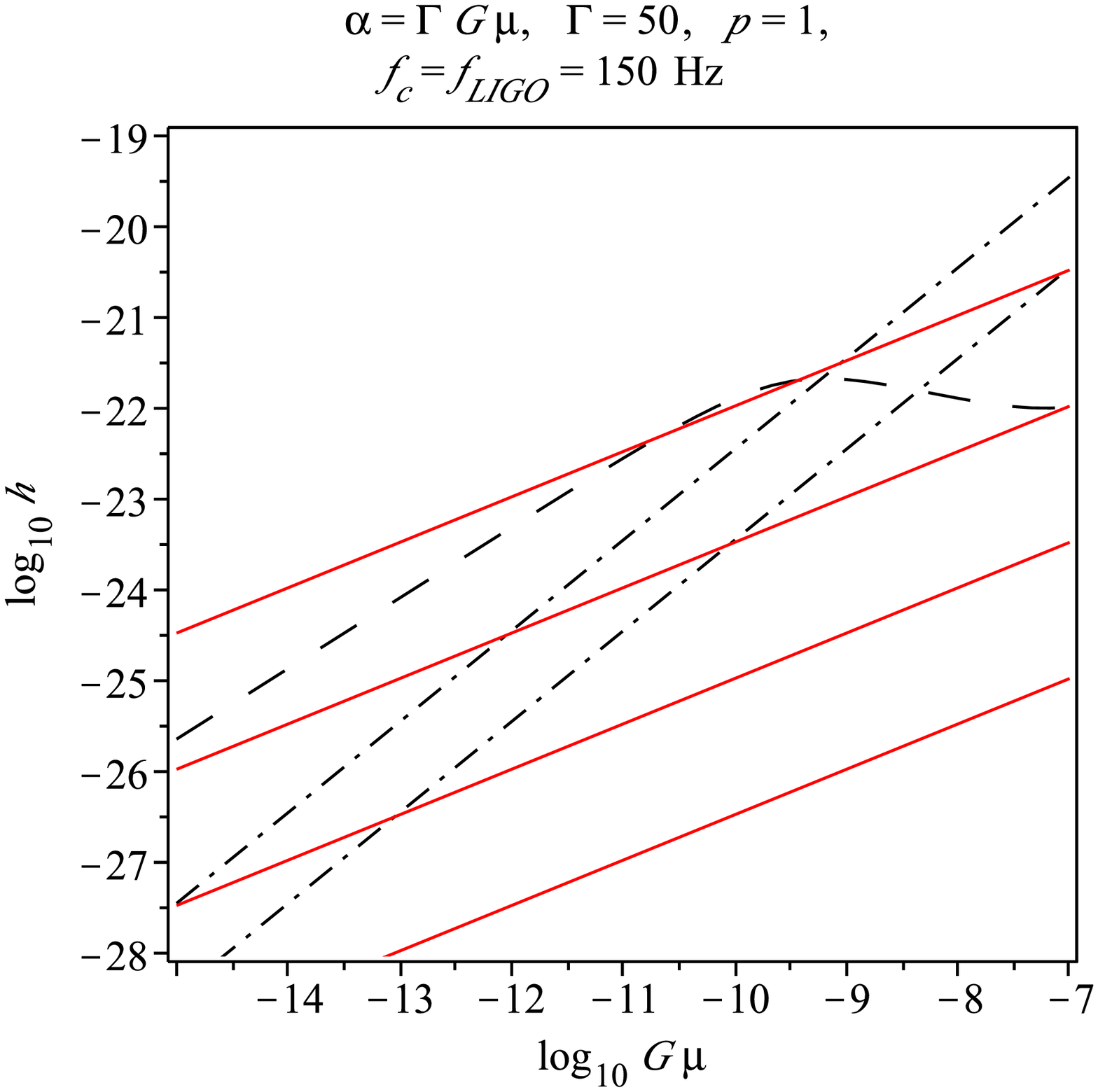} 
\end{minipage}
\begin{minipage}{0.48\textwidth}
\centering
  \includegraphics[scale=0.40]{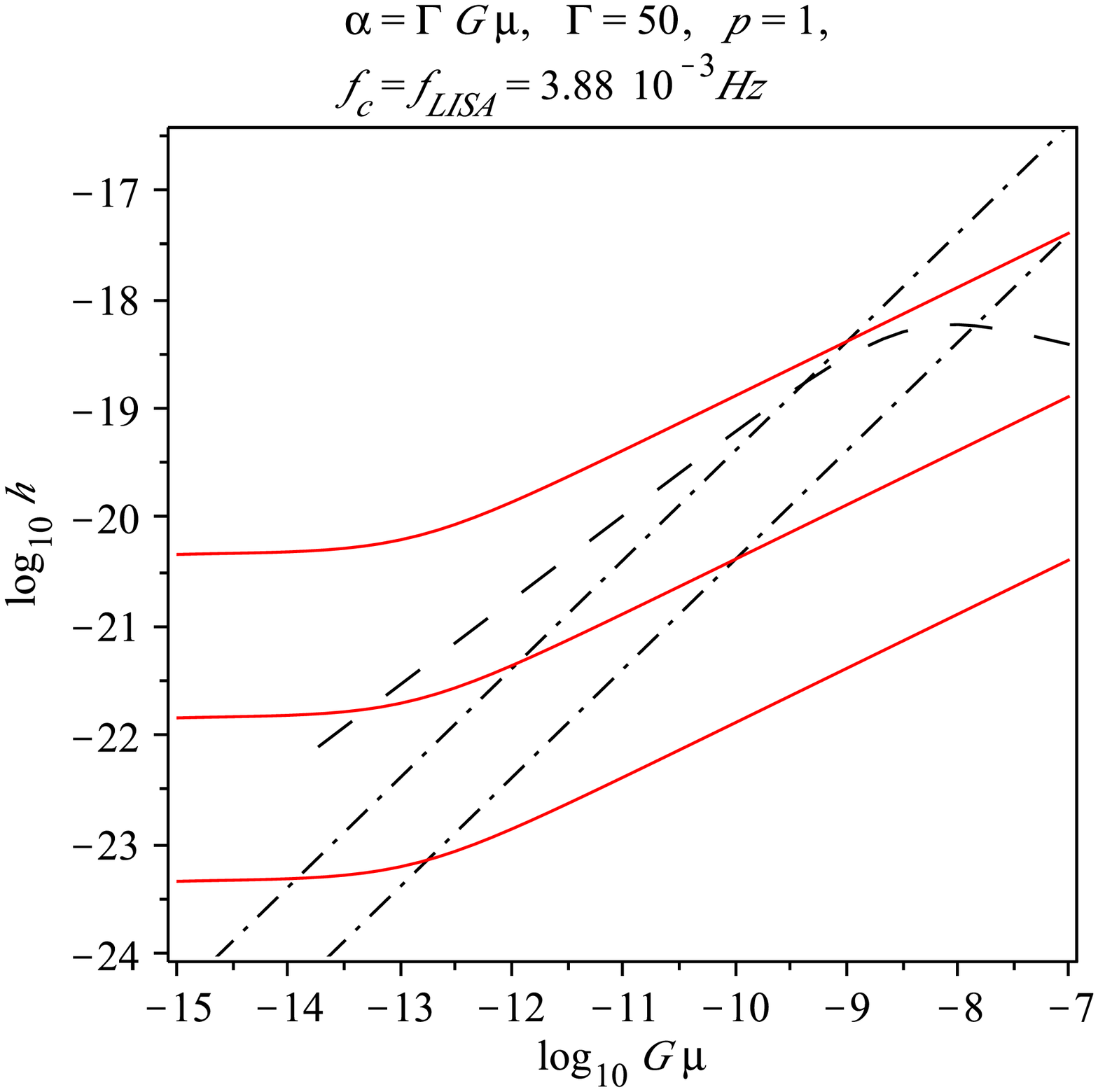} 
\end{minipage}
 \caption{The solid red lines give the amplitude of the stochastic gravitational wave background generated from the superposition of kink-kink bursts as a function of $G\mu$, in the frequency band of LIGO (left) and LISA (right) for different values of $qk'^2$ (from bottom to top: $qk'^2=1, 10^3, 10^6$ and also $10^9$ in the LIGO band). For comparison, we also plot the maximum amplitude of the individual bursts. The two parallel black dashed lines in both panels denote the cutoff amplitudes for kink bursts (for $q=1$ and $q=10^{-3}$) and the long-dashed black line is the amplitude of cusp bursts arriving at the instrument at the rate of 1/year. One sees that for large but reasonable values of $qk'^2$, the GW background hides the individual high frequency bursts.}
 \label{bback}
\end{figure}

We now compare the amplitude of the bursts with that of the dominant k-k contribution to the background in order to understand to what extent the latter acts as (self) confusion noise for the detection of individual bursts. In Fig \ref{bback} we plot both the amplitude of the k-k background for a range of different values of $qk'^2$ together with the maximal amplitudes of bursts, as a function of $G\mu$ and in the frequency bands of LIGO and LISA. The maximal amplitude for the bursts is taken to be the $q$-dependent cutoff amplitude given in Eq.~(\ref{conditionkinksjunctionsampl}) for the kink bursts, and the amplitude corresponding to a rate of 1 event per year for cusps.

Remarkably, the values of $qk'^2$ required for the background to dominate over the bursts are not particularly large, typically around $10^6$. \emph{Therefore there is a parameter regime in which kink proliferation on loops with junctions makes individual bursts simply unobservable.}

\section{Observational constraints on GW background}
\label{sec:Discussion}

We end with a discussion of the constraints on the different parameters of our cosmic-string model arising from observations of millisecond pulsars and LIGO, as well as the future prospects of observation with LISA. After a brief review of the current observations we summarize the resulting constraints in Section \ref{paramscontraintsalphaqkprimesquare} and discuss their implications for the GW burst signal in our model.

\subsection{Milisecond Pulsars}

The latest constraints on the dimensionless strain of gravitational waves imposed by millisecond pulsars are derived in \cite{Jenet:2006sv}. For a frequency dependence of the form
\begin{equation}
h_c(f)=A\left(\frac{f}{{\rm yr}^{-1}}\right)^{-3/2}
\label{fdependancepulsar}
\end{equation}
in the domain $[1/20~{\rm yr}^{-1},1~{\rm yr}^{-1}]$, the dimensionless amplitude $A$ is constrained to be
\begin{equation}
A<10^{-15} \, .
\label{pulsarconstraintamplitude}
\end{equation}
On comparison with the string spectrum calculated in  \eqref{hckkinterpolating}, it follows that \eqref{fdependancepulsar} can be used to apply a constraint provided the frequency interval $[(\alpha t_0)^{-1},z_{eq}(\alpha t_0)^{-1}]$ intersects 
$[\frac{1}{20}~{\rm yr}^{-1},1~{\rm yr}^{-1}]$. In practice, for the range of values of $\alpha$ we consider, this reduces to\footnote{The other condition, namely $z_{eq}(\alpha t_0)^{-1}>1/20~{\rm yr}^{-1}$ only imposes that $\alpha<10^{-5}$. We also note that the slope of the spectrum can never be $-1$ instead of $-3/2$ in this frequency interval for this would require values of $\alpha$ larger than $10^{-5}$.} $(\alpha t_0)^{-1}<1~{\rm yr}^{-1}$ {\it i.e.~}roughly $\alpha\gtrsim10^{-10}$.

Under this assumption and using Eqs.~\eqref{hckkinterpolating}-\eqref{pulsarconstraintamplitude}, we obtain the constraint $p^{-1/2} (qk'^2)^{1/2}<10^{-16} (1~{\rm yr})^{-3/2} t_0^{3/2} \Gamma$
which, since $p<1$ and taking $\Gamma=50$, yields the final order of magnitude constraint coming from pulsars
\begin{equation}
qk'^2<25 \qquad \qquad \text{ for } \alpha\gtrsim10^{-10}~i.e.~G\mu\gtrsim10^{-12}
\end{equation}
Note that within the range $\alpha>10^{-10}$, the constraint on parameter space is independent of $\alpha$.

\subsection{LIGO}

The ground based interferometer LIGO has already begun to take data. An upper limit on the amplitude of a stochastic background of gravitational waves based on the data from a two year science run, assuming a frequency dependence in the frequency band of LIGO ($\approx 10^2 {\rm Hz}$) of the form
\begin{equation}
\Omega_{gw}(f)=\Omega_\gamma \left(\frac{f}{100~Hz}\right)^\gamma \qquad \text{ around } 10^2~\text{Hz}
\end{equation}
is given in \cite{Abbott:2009ws}.
For the k-k background and for the range of values of $\alpha$ of interest, $\gamma=0$. Indeed, $h_c\propto f^{-1}$ for $f>\frac{z_{eq}}{\alpha t_0}$ and if $\alpha>10^{-16}$, $\frac{z_{eq}}{\alpha t_0}<10^2~{\rm Hz}$ so $\Omega_{gw}\propto f^2 h_c^2(f)\propto f^\gamma$ with $\gamma=0$ in the frequency band of LIGO.

The $95\%$ confidence upper limit on $\Omega_{gw}$ that applies to the k-k background reads 
\begin{equation}
\Omega_{gw}(f)<\Omega_0=6.9~10^{-6} \qquad \text{ around } 10^2\; {\rm Hz}
\label{LIGOomega}
\end{equation}
which, using \eqref{omegatohc} and $H_0=72$ km/s/Mpc translates into
\begin{equation}
h_c(f)<2.4~10^{-21}\left(\frac{1{\rm Hz}}{f}\right) \qquad \text{ around } 10^2\; {\rm Hz} \, .
\label{LIGOhc}
\end{equation}
This yields, in our parameter space, the constraint $p^{-1} qk'^2 \alpha<(2.4~10^{-21})^2\left(\frac{\Gamma}{10}\right)^2 t_0^2 z_{eq}$, i.e. using $\Gamma=50$ and $p<1$,
\begin{equation}
qk'^2 \alpha<0.1
\end{equation}
valid for
\begin{equation}
\alpha>10^{-16}.
\end{equation}

Note that the constraint on $\Omega_{gw}$ \eqref{LIGOomega} is expected to be improved by $4$ orders of magnitude with Advanced LIGO, which means an improvement by $2$ orders of magnitude on the constraint on $h_c$ \eqref{LIGOhc} and an improvement by $4$ orders of magnitude for the constraint on $\alpha q k'^2$.

\subsection{LISA}

\begin{figure}[H]
\centering
   \includegraphics[scale=1]{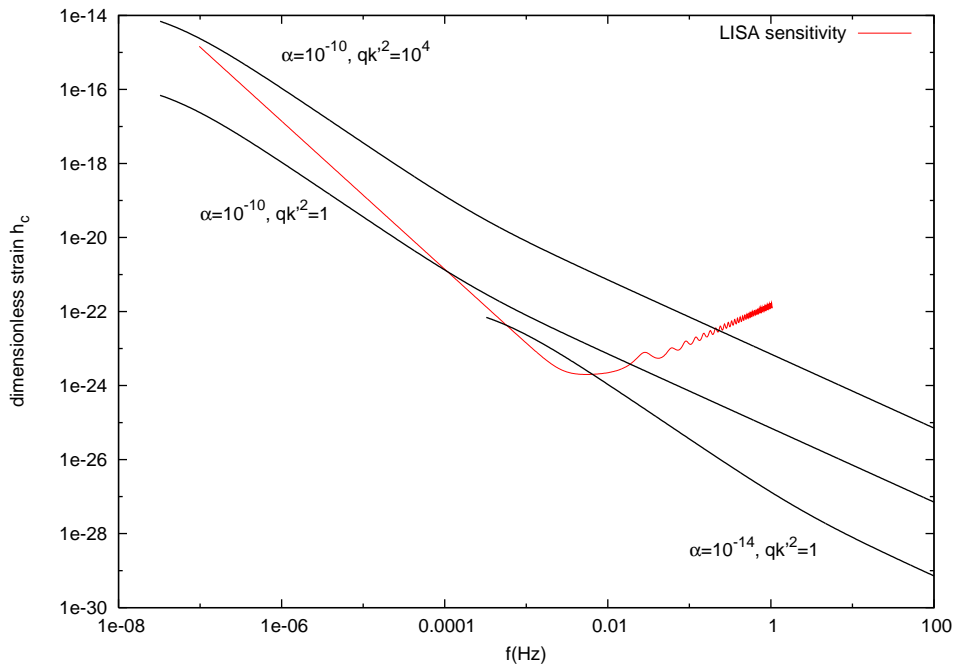} 
   \label{LISAbackground}
  \caption{The future space interferometer LISA should be able to detect the GW background generated from bursts emitted at kink-kink encounters for string tensions as low as $10^{-15}$ and for all values of $qk'^2>1$.}
\end{figure}

In the not-too-distant future, hopefully the space interferometer LISA will enable us to probe a different frequency band of gravitational waves around $\approx10^{-3}{\rm Hz}$. Using the sensitivity curves for the strain $h_c$ integrated over a year provided in \cite{LISAweb} with the default values of the parameters of LISA, one finds 
LISA should be able to detect the k-k background for $\alpha$ as small as  $10^{-14}$ even in models where $qk'^{2}$ is not much larger than $1$, as illustrated in Fig \ref{LISAbackground}.

\subsection{Discussion}
\label{discussion}

\begin{figure}[H]
\centering
   \includegraphics[scale=0.75]{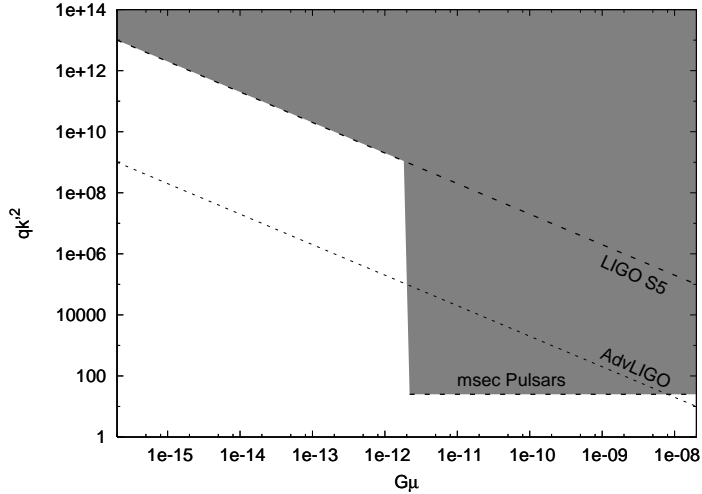} 
   \label{paramscontraintsalphaqkprimesquare}
  \caption{Summary of the different observational constraints on the parameters $G\mu$ and $qk'^2$. For tensions $G\mu \gtrsim 10^{-12}$ pulsar observations provide the most stringent constraints and enforce $qk'^2$ to be less than $\sim 10^2$. For smaller values of $G\mu$, LIGO places an upper bound on $qk'^2$ but this is very high. With Advanced LIGO one will be able to significantly improve on this. LISA in turn should be sensitive to all value of $(G\mu,qk'^2)$ in the range given here.}
\end{figure}

The observations of  millisecond pulsars and LIGO place rather stringent constraints on the different parameters of our cosmic-string model which are summarized in Fig \ref{paramscontraintsalphaqkprimesquare}, where we plot the various constraints in the $(G\mu,qk'^2)$ plane.

These observational constraints are relevant for our discussion in Section \ref{detect} of the GW burst signal in the class of models under consideration. Given the above assumptions about the network evolution and also \eqref{approx} and \eqref{timescaleassumptions}, we believe the most important astrophysical implications of this analysis are the following:

\begin{itemize}

\item For string tensions $G\mu\gtrsim10^{-12}$, the pulsar data imply $qk'^2 \lesssim10^2$. Hence models with an important kink proliferation on a significant proportion of the loops in the network are ruled out in this range of tensions. (Note that a similar conclusion applies to a standard network of cosmic strings with \emph{no junctions} and hence no kink proliferation. If the tension of these strings $G\mu\gtrsim10^{-12}$ then the number of kinks $k$ on the loops must satisfy $k\lesssim10$.)

As a consequence, in the remaining models in this range of tensions, the predictions of \cite{Damour:2001bk}  for the gravitational wave signal essentially apply. In particular, cusp bursts remain the dominant source of high frequency gravitational waves. Kink bursts should not be observable with LIGO or Advanced LIGO except for large tensions and provided $p$ is small. With LISA, bursts from kinks should be observable but with an SNR that is much smaller than the SNR of cusp bursts. The observation of kink bursts however should help to estimate the parameters values.

\item For string tensions $G\mu \lesssim 10^{-12}$, the current observational constraints on $qk'^2$ are much weaker and models exhibiting significant kink proliferation $(qk'^2\gg1)$ are allowed. If this is the case, the consequences on the GW predictions are the following. 

The SNR of kink bursts increases but only very slightly (never by more than one order of magnitude, see Figs.~\ref{courbesLIGO} and \ref{courbesLISAratep1}), because one quickly reaches the cutoff value given in \eqref{conditionkinksjunctionsampl}. In particular, individual kink bursts remain below the sensitivity of LIGO and Advanced LIGO. They will be accessible to LISA, with low SNR, only if $G\mu$ is close to $10^{-12}$ and $p$ is small. 

Furthermore, for large values of $qk'^2$ the GW signal from individual bursts in the LISA band is hidden by the confusion noise generated by smaller bursts from kink-kink encounters. Indeed, for values $qk'^2 \gtrsim10^4$, the GW background generated from the superposition of kink-kink bursts becomes so strong that it hides the individual sharp bursts from cusps and kinks (see Fig~\ref{bback}) .

Hence in this parameter regime the most important astrophysical implication of the presence of junctions is the absence of observable individual bursts and the presence of a strong GW background detectable by LISA and maybe by LIGO, with a spectrum given in \eqref{hckkinterpolating}.

\end{itemize}

\section{Conclusion}

A network of cosmic string loops generates a strongly non-Gaussian GW signal which includes sharp bursts at cusps and kinks that stand above the confusion noise generated by many smaller overlapping bursts.  For loops without junctions, the bursts at cusps provide the dominant contribution to the GW signal in all frequency bands of observational interest. However, in the presence of junctions the number of sharp kinks grows exponentially, leading to an  average number $k'$ of large amplitude kinks on loops with junctions that can be several orders of magnitude larger than that is expected on loops without junctions. In this paper we have studied the effect of this proliferation of kinks on cosmic string loops with junctions on the GW burst signal emitted from string networks of this kind.

We have calculated the rate of occurrence and the distribution in amplitude of the GW bursts emitted at cusps and kinks in the frequency bands of LIGO and LISA, as a function of the string tension $G\mu$, the number $k'$ of sharp kinks on loops with junctions, the fraction $q$ of loops that have junctions and the reconnection probability $p$. To do so we have used a simple `scaling' model for the cosmic string network. Our predictions for GW observations depend on the simplifying assumptions involved in this model. These include, in particular, that loops are characterized by a single length scale $L$ (see Eq.~\eqref{approx}), and that all loops formed at time $t$ have the same size $L \sim \alpha t$, with $\alpha = 50 G \mu$.

The amplitude of individual kink bursts depends on the combination $qk'$. In models with junctions one generally expects $qk' \gg 1$, whereas in models without junctions one usually assumes loops have only one or a few kinks. We have seen that this difference lowers the redshifts required to achieve observable rates and therefore can lead to a significant enhancement of the GW burst signal of kinks in loops with junctions relative to standard loops for given values of the string tension and $p$. 

In fact, one might have thought that by taking $qk'$ sufficiently large, one could easily make the SNR level of kink bursts much larger than the SNR of cusps. However, we have seen there is an important limitation that arises from the density of loops in the model. This is because below a certain redshift the probability of finding a loop -- and therefore a GW burst-- becomes exceedingly small. Since the largest amplitude bursts are emitted at the smallest redshift this translates, in the scaling model, in an upper bound on the amplitude of kink bursts. The latter depends on $G\mu$, $p$ and $q$ but lies, for string tensions below $10^{-10}$, always below the amplitude of cusp bursts both in the LIGO and LISA frequency bands. For larger string tensions this is a priori not the case. 
However, in this case significant kink proliferation is ruled out by millisecond pulsar observations, so that cusps remain the strongest source of gravitational wave bursts in both bands.

Another effect of kink proliferation is to significantly enhance the stochastic background generated by the superposition of overlapping bursts at the detector. In Section \ref{stoch}, we computed analytically the spectrum of the kink and of the kink-kink backgrounds. We showed that as soon as $qk'^2\gtrsim1$, the latter becomes dominant over the cusp and the kink backgrounds.

By combining our results on the stochastic background with the latest observations from millisecond pulsars and LIGO, we were able to constrain the parameters in the model. 
For string tensions $G\mu\gtrsim10^{-12}$, we find significant kink proliferation is not allowed and hence as far as the GW predictions are concerned, the results of \cite{Damour:2001bk} essentially apply. For $G\mu\lesssim10^{-12}$ the observational constraints on kink proliferation are much weaker. In models with a large number of kinks on loops, the SNR of individual kink bursts in LISA or LIGO increases though only by one order of magnitude at most because of the cutoff mentioned above. Most importantly, however, the stochastic background generated from the incoherent superposition of kink-kink bursts is strongly enhanced and becomes dominant over the cusp and kink backgrounds when $qk'^2\gtrsim1$, and it even hides all individual bursts from both cusps and kinks for $qk'^2\gtrsim10^4$.

\section*{Acknowledgements}

We thank Ed Porter and Christophe Ringeval for useful comments and discussions. DAS is grateful to the CERN theory group for hospitality whilst this work was in progress. This work was supported by the Agence Nationale de la Recherche grant "STR-COSMO" (ANR-09-BLAN-0157).

\appendix
\section{GW signal dominated by large amplitude kinks}
\label{appendix:smallkinks}

The goal of this appendix is to illustrate the fact that GWs produced by kinks on loops with junctions are dominated by large amplitude kinks ($A\gsim 0.25$), despite the larger number of  small amplitude kinks ($A\ll1$). (Recall that whenever a kink propagates through a junction it generates 3 daughter kinks whose amplitude, is generally smaller than that of the initial kink \cite{Binetruy:2010bq}.)  As a result, when computing the bursts rates or the characteristic amplitude of the stochastic background produced by the incoherent superposition of many bursts, we will not need to use the full distribution of kink amplitudes on such loops but instead only the number $k'$ of kinks that have an amplitude $\gsim 0.25$ as done throughout the paper.

Let $f(A)$ be the distribution of kink amplitudes on a loop at the end of the proliferation phase (so that $f(A)dA$ is the number of kinks that have an amplitude between $A$ and $A+dA$). In principle, $f(A)$ depends on the initial conditions of the particular loop we are considering. But, as was argued in \cite{Binetruy:2010bq}, the details of these initial conditions are lost after a few time intervals of $L$ (the typical size of the loop). We will therefore consider $f(A)$ to be the same for all the loops in the network. 

The form of $f(A)$ was computed numerically in \cite{Binetruy:2010bq} for an idealized loop, using a tree construction. Apart from the string tensions, the main parameter in that setup is the number $n$ of generations for which proliferation takes place (physically its time duration in units of $L$). Simulations of realistic loops lead us to consider $n$ of the order of $10$ but this parameter could be much larger. From now on we will use the notation $f_n(A)$ when we want to indicate the particular value of $n$ that we used, and simply $f(A)$ elsewhere.

The total number of kinks obviously increases (exponentially) with $n$ but $f_n(A)$ becomes peaked around some small value that goes to zero exponentially fast with $n$: the vast majority of kinks have a very small amplitude. However, the number of kinks that have an amplitude larger than some fixed value also increases exponentially.

\subsection{Burst rates}
Intuitively, small amplitude kinks produce bursts that can only be observed experimentally if they were produced at small enough redshift. The question of their negligeability is therefore a competition between their large number and the fact that the number of loops decreases with redshift.

To be more precise, let us determine the rate of events $\dot{N}(h)$ that have an amplitude larger than some fixed value $h$ (that could be for instance the experimental threshold).

To answer this question, we need to slightly modify formula \eqref{obssignal} to take into account the fact that we no longer consider kinks of amplitude $A=1$ exclusively. The amplitude of a GW burst produced at a kink of amplitude $A$ on a loop at redshift $z$ (and observed at some frequency $f$) is simply
\begin{equation}
\label{obssignalwithamp}
h(A,z) = A \frac{G \mu L}{((1+z) Lf)^{2/3}}\frac{1+z}{t_0 z}
\end{equation}
where we have now removed the $\Theta$ function that appeared in \eqref{obssignal} because for reasonable values of $f$ (in the band of LISA or LIGO for instance) it only imposes a higher cutoff value on $z$ that is much larger than all redshifts of interest for bursts. We will also need an updated version of \eqref{diffratekinks} to compute the rate of bursts. We now define $d\dot N(A,z)$ the number of bursts reaching us and that were emitted between $z$ and $z+dz$ from kinks with an amplitude between $A$ and $A+dA$. This quantity reads
\begin{equation}
\label{diffratekinkswithamp}
d\dot N(A,z)=\frac{\theta_m}{L} (1+z)^{-1} q n(z) f(A) dV(z) dA
\end{equation}

We can simplify the expressions above by remarking that we are only interested in bursts coming from the region $z\ll1$ (see section 3 of this paper). In this region, the density of loops $n(z)$, the beaming angle $\theta_m$ and the typical size of the loops $L$ do not depend on $z$ and can be approximated by their values today. We could express those in terms of our parameters $p$, $\alpha$ but we are only interested here in the form of the dependence on $A$ and $z$ so we will simply use the notations $n_0$, $\theta_0$ and $L_0$. Furthermore, we will use $dV(z)\approx t_0^3 z^2 dz$. Defining $C= G \mu {L_0}^{1/3}f^{-2/3} t_0^{-1}$ and $D=\theta_0 L_0^{-1} q n_0$, we obtain

\begin{equation}
\label{withampsimplified}
h(A,z) = C \frac{A}{z} \qquad \text{ and } \qquad d\dot N(A,z)=D f(A) z^2 dz dA
\end{equation}

It is then clear that $\dot{N}(h)$ is obtained by integrating $d\dot N(A,z)$ on the domain $\mathcal{D}(h)$ shown below.

\begin{figure}[H]
 \centering
   \includegraphics[scale=0.60]{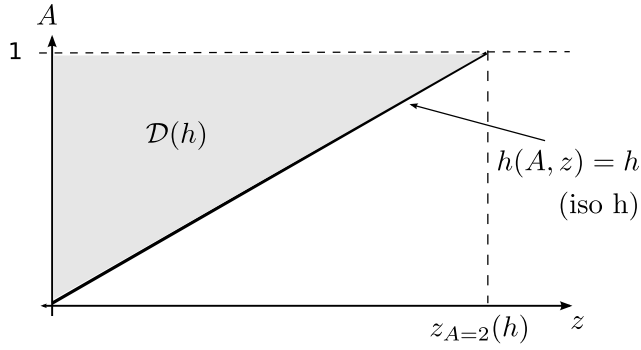} 
   \label{domainintegrationbursts}
 \caption{Domain of integration of $d\dot N(A,z)$ required to compute $\dot{N}(h)$}
\end{figure}

\begin{equation}
\dot{N}(h)=\int_{\mathcal{D}(h)}d\dot N(A,z)=D\int_{A=0}^{1}f(A)\left(\int_{z=0}^{A\frac{C}{h}}z^2dz\right)dA=\frac{D C^3}{3h^3}\int_{A=0}^{1}f(A)A^3dA
\label{burstintegralonA}
\end{equation}

The question is now to know if this last integral is dominated by large values of $A$ or by the $A\ll1$ region. We define for convenience the funtions
\begin{equation}
F_{k,n}(x)=\int_x ^1 f_n(A) A^k dA
\end{equation}

\begin{figure}[H]
 \centering
   \includegraphics[scale=0.80]{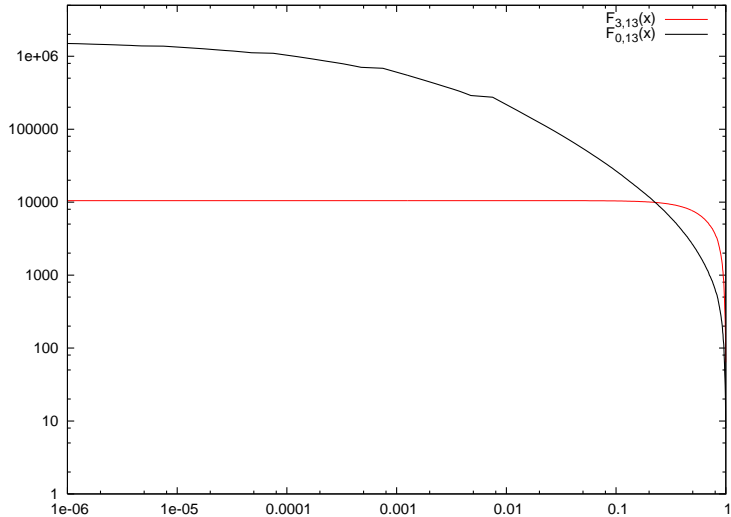} 
   \label{burstdomination}
 \caption{As $x$ decreases from $1$ to $0$, $F_{3,13}(x)$ (red) strongly grows and quickly reaches a plateau which shows that the main contribution to the integral appearing in \eqref{burstintegralonA} (namely $F_{3,13}(0)$) comes from $A\approx1$. By contrast, the main contribution to the kink distribution (illustrated by $F_{0,13}$) comes from smaller amplitudes. A good approximation to the integral in \eqref{burstintegralonA} is $F_{3,13}(0)\approx F_{0,13}(1/4)$}
\end{figure}

We see from figure \ref{burstdomination} that the dominant contribution to $\int_{A=0}^{1}f(A)A^3dA$ comes from $A\approx1$ (as $x$ starts decreasing from $1$, the red curve quickly reaches a plateau).  In that figure, we used $n=13$.

Since the dominant contribution comes from $A\approx 1$, we can estimate the integral by replacing $A^3$ by $1$ and simply counting the number $k'$ of kinks that have an amplitude of the order of $1$, i.e. between $1$ and say $1/4$
\begin{equation}
\int_{A=0}^{1}f(A)A^3dA\approx \int_{A=1/4}^{1}f(A)A^3dA\approx 1^3 \int_{A=1/4}^{1}f(A)dA=k'
\end{equation}

We see from figure \ref{burstdomination} that such an estimation is a very good approximation at least for our order of magnitude calculations: the intersection between both curves occurs close to $x=1/4$ and the red one has already reached its plateau value. We checked that this was not very sensible to a different choice of $n$. Of course, the value of $k'$ does depend on the number of generations $n$ and we leave it as a free parameter.\\

Therefore, only large amplitude bursts contribute to the rate of bursts we might observe and the picture that we have used in this paper, namely loops with $k'$ kinks that all have the same amplitude $A=1$ yields a very good approximation for the rate of bursts.

\subsection{Stochastic background}

We now address the question of the importance of small amplitude kinks in the background produced by the incoherent superposition of many bursts? The (squared) characteristic amplitude of such a background is computed by summing the squared amplitude of bursts that actually overlap at the detector.

\begin{figure}[H]
 \centering
   \includegraphics[scale=0.60]{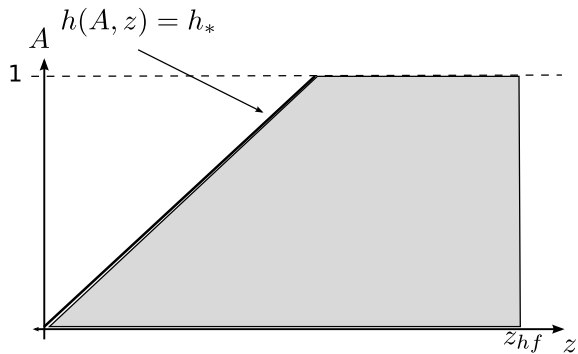} 
   \label{domainintegrationbackground}
 \caption{Domain of integration of $f^{-1}h^2(A,z)d\dot{N}(A,z)$ required to compute $h_c$.}
\end{figure}

In situations (as the one we consider throughout this paper except for this appendix) where there is a direct correspondance between the amplitude of the bursts and the redshift (and also between rates and redshifts), this sum can be expressed as in \eqref{intbackground} as an integral over redshifts larger than the value $z_{b\rightarrow b}$ for which the rate becomes larger than the inverse timescale of the experiment. If we allow for a whole distribution of kink amplitudes instead of considering that all kinks have amplitude $1$, the situation becomes more complicated. We use the same procedure as \cite{Siemens:2006yp} to remove the infrequent (non overlapping) large amplitude bursts, namely we only include in the sum bursts that have an amplitude $h$ smaller than the value $h_*$ for which $\dot{N}(h)>f$. This amounts to performing the integral of the quantity $f^{-1}h^2(A,z)d\dot{N}(A,z)$ over the domain $\mathcal{D'}$ showed in figure \ref{domainintegrationbackground}.

\begin{figure}[H]
 \centering
   \includegraphics[scale=0.80]{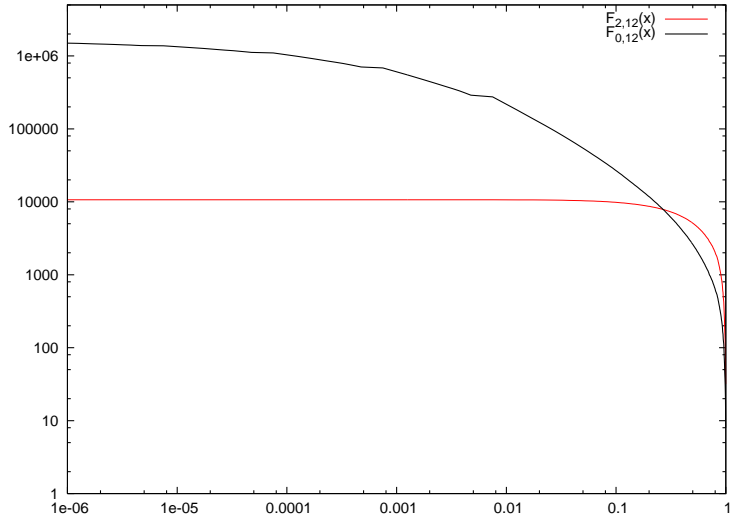} 
   \label{backgrounddomination}
 \caption{As $x$ decreases from $1$ to $0$, $F_{2,13}(x)$ (red) strongly grows and quickly reaches a plateau which shows that the main contribution to $\int_{A=0}^1 f(A) A^2 dA$ comes from $A\approx1$. A good approximation to $\int_{A=0}^1 f(A) A^2 dA$ is $F_{0,13}(1/4)$ i.e. the number of large amplitude kinks that we studied in \cite{Binetruy:2010bq}}
\end{figure}

Let $z(A)$ be defined by $h(A,z(A))=h_*$. If $z(A=1)\ll1$ (which is the most frequent case) then this is a stright line in the (A,z) plane. The integral can be written
\begin{equation}
h_c^2=\int_{A=0}^1 f(A) A^2\left(\int_{z(A)}^{z_{hf}} \frac{d\dot{N}(A=1,z)}{f} h^2(A=1,z) dz\right)dA
\end{equation}
Except for its bounds (and obviously for a factor $k'$), the inside integral over redshifts turns out to be exactly the one that we encountered for the calculation of $h_c$ in \ref{intbackground}. We already showed that it was never sensible to the variations of its lower bound. We are therefore simply left with $\int_{A=0}^2 f(A) A^2 dA$ as an overall multiplicative factor. As in the case described above of bursts, this integral is dominated by $A\approx1$ and an be approximated by $k'=\int_{1/4}^1 f(A)$ the number of kinks of large amplitude. The validity of this approximation is illustrated in figure \ref{backgrounddomination}

Here again, simply considering that loops have $k'$ large amplitude kinks instead of considering a whole distribution of kink amplitudes yields the correct result for the background.

\vskip 2cm

\section{Redshift integrals}
\label{appendix:integrals}
We provide an analytical description of the integrals that appear in the calculation of the different types of backgrounds. We will always assume here that $z_a\ll z_b$.

\subsection{The kink-kink integral}
\begin{figure}[H]
\centering
   \includegraphics[scale=0.40]{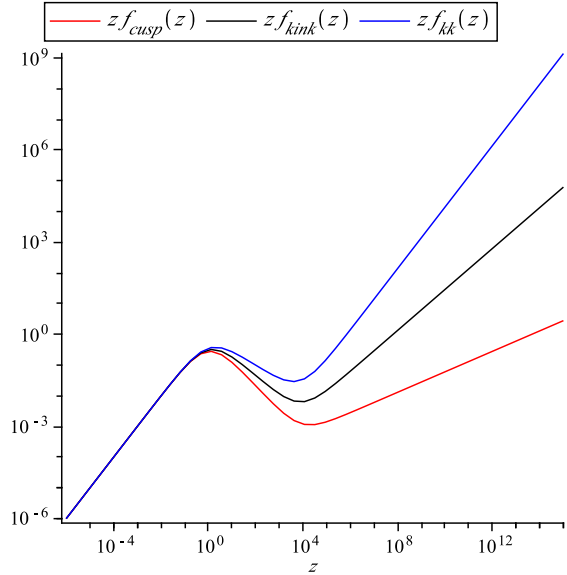} 
   \label{integralappendixall}
  \caption{In the case of the kink or the kink-kink backgrounds, the integral is either dominated by large redshifts or by $z\approx1$. The situation is very different for cusps where the domination comes from the smallest redshifts (or from $z\approx 1$)}
\end{figure}

In this appendix, we study the behaviour of the function
\begin{equation}
g_{kk}=(z_a,z_b)\rightarrow \int _{z_a} ^{z_b} \left(\frac{1+z/z_{eq}}{1+z}\right)^{3/2}dz=\int _{z_a} ^{z_b} f_{kk}(z) dz
\end{equation}
In order to give an idea of where the main contributions to this integral come from, we plot the function $z f_{kk}(z)$ in figure \ref{integralappendixall}.

This function has a local maximum around $z=1$, a local minimum around $z=z_{eq}$ and then diverges (as $\frac{z}{ z_{eq}^{3/2}}$) when $z$ goes to infinity. The value at which it becomes larger than the local maximum is $z\approx z_{eq}^{3/2}$.

Using the fact that for $z\ll1$, $f_{kk}(z)\approx1$, for $1\ll z\ll z_{eq}$, $f_{kk}(z)\approx z^{-3/2}$ and for $z\gg z_{eq}$, $f_{kk}(z)\approx z_{eq}^{-3/2}$, it is easy to see that $\int_0^{z_{eq}^{3/2}}f(z)dz=\mathcal{O}(1)$ and that the dominant contribution to this integral comes from an interval around $z=1$. Therefore, in our two regimes of interest\\

\begin{itemize}
\item $z_a\ll1$ and $1\ll z_b\ll z_{eq}^{3/2}$: then $g$ doesn't depend (at leading order) on $z_a$ or $z_b$: $g(z_a,z_b)\approx 1$\\

\item $z_a\ll z_{eq}^{3/2}$ and $z_b\gg z_{eq}^{3/2}$: then $\int_{z_{eq}^{3/2}}^{z_b}f(z)dz\approx \frac{z_b}{z_{eq}^{3/2}}\gg1$ dominates over the rest of the integral and $g(z_a,z_b)\approx \frac{z_b}{z_{eq}^{3/2}}$.

\end{itemize}

\subsection{The kink integral}
The integral appearing in the calculation of the kink background is
\begin{equation}
g_{kink}=(z_a,z_b)\rightarrow \int _{z_a} ^{z_b} \frac{(1+z/z_{eq})^{4/3}}{(1+z)^{5/3}}dz=\int _{z_a} ^{z_b} f_{kink}(z) dz
\end{equation}
As can be seen from figure \ref{integralappendixall}, the shape of $z f_{kink}(z)$ is the same as in the kk case so the discussion above applies after correctly changing the exponents. In particular, for our regimes of interest
\begin{itemize}
\item $z_a\ll1$ and $1\ll z_b\ll z_{eq}^{2}$: $g(z_a,z_b)\approx 1$\\

\item $z_a\ll z_{eq}^{2}$ and $z_b\gg z_{eq}^{2}$:  $g(z_a,z_b)\approx \frac{z_b^{2/3}}{z_{eq}^{4/3}}$.

\end{itemize}

\subsection{The cusp integral}
The integral appearing in the calculation of the cusp background is
\begin{equation}
g_{cusp}=(z_a,z_b)\rightarrow \int _{z_a} ^{z_b} \frac{(1+z/z_{eq})^{7/6}}{(1+z)^{-11/6}}dz=\int _{z_a} ^{z_b} f_{cusp}(z) dz
\end{equation}
As before, in an effort to identify the main contributions to such an integral, we plot $z f_{cusp}(z)$
The situation is very different than in the kk case. There is again a local maximum around $z=1$ where $z f_{cusp}(z)=1$ and then we have $z f_{cusp}>1$ for $z>z_{eq}^{7/2}$. Among the regimes of interest, there will be one where both $z_a$ and $z_b$ lie in the interval $[1,z_{eq}^{7/2}]$ which explains why the value of $z_a$ is crucial in the case of cusps.

\bibliographystyle{utphys}
\bibliography{bibliofinal}

\end{document}